\documentclass[12pt]{article}
\usepackage{enumerate}
\usepackage{natbib}
\usepackage{amsmath,amssymb}
\usepackage{bbold}
\usepackage{graphicx,graphics}
\usepackage{caption}
\usepackage{multirow}
\usepackage{booktabs}
\usepackage{caption}
\usepackage{algorithm}
\usepackage{algpseudocode}
\usepackage{comment}
\usepackage{color}
\usepackage{cancel}
\usepackage{mathdots}
\usepackage{bbm}
\usepackage{subfig}
\usepackage[export]{adjustbox}
\usepackage{enumitem}
\usepackage{url} 

\usepackage{setspace}

\newcommand{\blind}{1}
\graphicspath{{Figs/}}
\def\argmax{\mathop{\rm argmax}}
\def\argmin{\mathop{\rm argmin}}
\def\arginf{\mathop{\rm arginf}}

\newcommand{\real}{\ensuremath{\mathbb{R}}}

\newcommand{\ltwo}{\ensuremath{\mathbb{L}^2}}

\newcommand{\inner}[2]{\left\langle #1,#2 \right\rangle}

\makeatletter
\newenvironment{breakablealgorithm}
  {
   \begin{center}
     \refstepcounter{algorithm}
     \hrule height.8pt depth0pt \kern2pt
     \renewcommand{\caption}[2][\relax]{
       {\raggedright\textbf{\fname@algorithm~\thealgorithm} ##2\par}%
       \ifx\relax##1\relax 
         \addcontentsline{loa}{algorithm}{\protect\numberline{\thealgorithm}##2}%
       \else 
         \addcontentsline{loa}{algorithm}{\protect\numberline{\thealgorithm}##1}%
       \fi
       \kern2pt\hrule\kern2pt
     }
  }{
     \kern2pt\hrule\relax
   \end{center}
  }
\makeatother

\begin{document}

\def\spacingset#1{\renewcommand{\baselinestretch}%
{#1}\small\normalsize} \spacingset{1}


\if1\blind
{
  \title{\bf Scalar-on-Shape Regression Models for Functional Data Analysis}
  \author{Sayan Bhadra and Anuj Srivastava\thanks{
    The authors gratefully acknowledge {NSF DMS 1953087, NSF, DMS 2413748, NIH R01 MH120299.  }}\hspace{.2cm}\\
    Department of Statistics, Florida State University}
  \maketitle
} \fi

\if0\blind
{
  \bigskip
  \bigskip
  \bigskip
  \begin{center}
    {\LARGE\bf Scalar-on-Shape Regression Models for Functional Data Analysis}
\end{center}
  \medskip
} \fi

\bigskip
\begin{abstract}
Functional data contains two components:  shape (or amplitude) and phase. This paper focuses on a branch of functional data analysis (FDA), namely Shape-Based FDA, that isolates and focuses on shapes of functions. Specifically, this paper focuses on Scalar-on-Shape (ScoSh) regression models that incorporate the shapes of predictor functions and discard their phases. This aspect sets ScoSh models apart from the traditional Scalar-on-Function (ScoF) regression models that incorporate full predictor functions. ScoSh is motivated by object data analysis, {\it, e.g.}, for neuro-anatomical objects, where object morphologies are relevant and their parameterizations are arbitrary. 
ScoSh also differs from methods that arbitrarily pre-register data and uses it in subsequent analysis. In contrast, ScoSh models perform registration during regression, using the (non-parametric) Fisher-Rao inner product and nonlinear index functions to capture complex predictor-response relationships. This formulation results in novel concepts of {\it regression phase} and {\it regression mean} of functions. Regression phases are time-warpings of predictor functions that optimize prediction errors, and regression means are optimal regression coefficients. 
We demonstrate practical applications of the ScoSh model using extensive simulated and real-data examples, including predicting COVID outcomes when daily rate curves are predictors. 
\end{abstract}

\noindent%
{\it Keywords:}  shape regression, shape models, COVID data analysis, functional shapes, shape-based FDA, functional regression analysis. 

\spacingset{1.9} 
\section{Introduction and Literature Survey}
\label{sec:intro}
    Rapid advances in data collection and storage technologies have led to a surge in problems where the data objects are functions recorded over time and space. Functional datasets in neuroimaging, biology, epidemiology, meteorology, and finance have fuelled a growing interest in Functional Data Analysis (FDA). FDA deals with statistical analysis, including clustering, summarizing, modeling, and testing functional data. Functional regression incorporates functional variables in regression models as predictors, responses, or both. Specifically, {\it Scalar-on-Function} (ScoF) regression occurs when the predictors are functions and responses are scalar (or vectors). This problem has widespread applications in many scientific domains, with several examples presented later in this paper. Scalar-on-function regression is a natural extension of the standard multivariate regression model for the FDA. 

    Our focus differs from traditional ScoF by emphasizing the \textit{shapes} (amplitudes) of functions rather than the functions themselves. This focus is motivated, for example, by problems in neuroimaging where morphologies of anatomical objects are used to predict clinical measurements. Accordingly, we develop a regression model where \textit{shapes of functions} are predictors for scalar responses. Mathematically, \textit{shape} is a property that is invariant to certain transformations considered nuisances in shape analysis~(\cite{kendall-barden-carne,dryden2016statistical}). For scalar functions, $\{f_i\}$, shapes relate to the number and height of extremes (peaks and valleys), but the locations are considered nuisances. Changes in the locations of these points, represented by diffeomorphisms $\{\gamma_i\}$ and implemented using the transformation $\{f_i \mapsto f_i \circ \gamma_i\}$, are called phase changes and are ignored in shape analysis. Thus, the shapes of a function $f_i$ and its diffeomorphic time warping $f_i \circ \gamma_i$ are deemed identical. Shape-based FDA (see~\cite{wu-TEST:2023, SK2016,marron-etal-EJS:2014,marron-etal-Statsc:2015,stoecker-etal:2023}) is gaining interest, especially when phase variability is less critical, such as in COVID-19 rate curves where peaks represent pandemic waves. Several methods ({\it e.g.}, ~\cite{marron-etal-EJS:2014,marron-etal-Statsc:2015, SK2016}) have been developed to separate shape from phase as a stand-alone tool in FDA. This paper introduces a regression model that separates phases and amplitudes within the statistical model, not as a preprocessing step. This approach can enhance performance and interpretation by optimizing the phases when they are uninformative. The literature on shape regression is scarce, but extensive research exists in some related areas. We summarize these contributions next. 

    \begin{itemize}[leftmargin=*]
    \item {\bf Scalar-on-Function (ScoF) Regression Models}:  We start with the basic (parametric) functional linear model of (FLM) of~\cite{RS2005}: 
    \begin{equation}
    y_i=\alpha + \inner{\beta}{f_i} +\epsilon_i,\ i=1,2,\dots, n,
    \label{eq:classic-FLM}
    \end{equation}
  where $f_i: [0,T] \to \real$ is the predictor (element of a function space ${\cal F}$) and $y_i \in \real$ is the response. Also, $\alpha \in \real$ is the offset, $\beta \in {\cal F}$ is the coefficient function, and $\epsilon_i \in \real$ are the measurement errors. Here $\inner{\beta}{f_i} = \int \beta(t) f_i(t)~dt$ denotes the $\ltwo$ inner product. FLM assumes  {\it i.i.d} observation noise, $\epsilon_i \sim\mathcal{N}(0,\sigma^2)$.  To estimate $\beta$, one commonly minimizes the term $\sum_{i=1}^n (y_i- \alpha - \inner{\beta}{f_i})^2$. 
  ~\cite{R2012} used principal components of the predictor functions as an orthonormal basis for $\beta$. 
  To regularize $\beta$, one adds a penalty term $\lambda {\cal R}(\beta)$,  where $\lambda>0$ is a tuning parameter; see \cite{ME1999,J2009,RO2007,LP2011,Z2012}.


\cite{AS2008} introduced non-linearity to regression models by introducing a function $h: \real \to \real$ to result in the model $y_i=h\left(\inner{\beta}{f_i}\right)+\epsilon_i$. They used a kernel to estimate $h$, whereas \cite{E2009} alternatively optimize $\beta$ and $h$ with smoothness constraints.
Several authors (\cite{JS2005,A2006,F2013}) have studied multiple index models of the type: 
    \begin{equation} \label{eq:multi-index-flm}
    y_i=\alpha_0+\sum_{j=1}^r h_j \left(\inner{\beta_j}{f_i}\right)+\epsilon_i \ ,
    \end{equation}
for an arbitrary $r$. 
~\cite{M2013} further generalized the model using a time-indexed set of functions: $y_i=\int_0^1 \inner{\beta_{t, \cdot}}{f_i(\cdot)}~dt + \epsilon_i$,
(where $\beta$ is now a bi-variate function). Amongst notable nonparametric approaches, \cite{boj2010distance} introduced a weighted distance-based regression for functional predictors using semi-metrics on function spaces. 
\cite{boj2016global} introduced non-parametric link functions to generalize their earlier models.

\item {\bf Shape-on-Scalar (ShoSc) Regression Models}: There is extensive literature on the inverse problem, where the shapes of functions form responses, and predictors are Euclidean vectors, see ~\cite{lin2017extrinsic,shi2009intrinsic,tsagkrasoulis2018random}. An example of this problem is when the scalar predictor is time, and the goal is to fit a time curve on a shape space given finite observations. This also relates to fitting {\it smoothing splines} on shapes. Intrinsic manifold valued regression models have been studied widely by \cite{ghosal2023}, \cite{petersen2019frechet}, whereas extrinsic models have been studied by \cite{lin2019extrinsic}.  A wide literature on geodesic regression (\cite{thomas2013geodesic,shin2022robust})
also belongs to this category.  

\begin{table}
\caption{Listing of various models studied in this paper} \label{tab:list-models}
\begin{tabular}{|ll|}
        \hline 
{\small\bf Single-Index Scalar-on-Shape (SI-ScoSh)} & \hspace{-0.8cm} $y_i=g(f_i(0))+h\left(\sup_{\gamma_i\in\Gamma}{\left\langle\beta,q_i\star\gamma_i\right\rangle}\right)+\epsilon_i$ \\
\hline 
{\small\bf Single-Index Scalar-on-Function (SI-ScoF)($\ltwo$)} &    \hspace{-0.2cm}  $y_i=c_i+h\left({\left\langle\beta,f_i\right\rangle}\right)+\epsilon_i$ \\
\hline 
{\small \bf Single-Index Scalar-on-Function (SI-ScoF)($FR$)} &
\hspace{-0.2cm} $y_i=g(f_i(0))+h\left({\left\langle\beta,q_i\right\rangle}\right)+\epsilon_i$ \\
\hline 
{\small\bf Scalar-on-Shape (ScoSh)}: & \hspace{-0.8cm}  $y_i=g(f_i(0))+\sup_{\gamma_i\in\Gamma}{\left\langle\beta,q_i\star\gamma_i\right\rangle}+\epsilon_i$\\
\hline 
{\small\bf Scalar-on-Function (ScoF)($\ltwo$)}: & 
\hspace{-0.8cm} $y_i=g(f_i(0))+{\left\langle\beta,f_i\right\rangle}+\epsilon_i$\\
\hline 
{\small\bf Scalar-on-Function (ScoF)($FR$)}: & 
\hspace{-0.8cm} $y_i=g(f_i(0))+{\left\langle\beta,q_i\right\rangle}+\epsilon_i$\\     
\hline
\end{tabular}
\end{table}
\item {\bf Scalar-on-Shape (ScoSh) Regression Models}: 
\cite{ahn-CSDA} first studied a ScoSh model but with a major limitation. Since $y_i$s depend on the shapes of $f_i$s, they must be invariant to phase changes in $f_i$s. Thus, the response should remain unchanged if $f_i$ is replaced by $f_i \circ \gamma$ in the model. In Eqns.~\ref{eq:classic-FLM} and \ref{eq:multi-index-flm}, the $\ltwo$ inner-product fails to provide this invariance because $\inner{\beta}{f_i} \neq \inner{\beta}{f_i \circ \gamma_i}$ generally. Even under identical dual transformation, we don't have equality, {\it i.e.},  $\inner{\beta}{f_i} \neq \inner{\beta\circ \gamma}{f_i \circ \gamma}$. This rules out using $\sup_{\gamma} \inner{\beta}{f_i \circ \gamma}$ to remove phase variability, as it is degenerate and not phase-invariant (see~\cite{SK2016}). \cite{ahn-CSDA} replaced the $\ltwo$ inner-product $\inner{\beta}{f_i}$ in Eqns \ref{eq:classic-FLM}, \ref{eq:multi-index-flm} with $\sup_{\gamma_i} \inner{\beta}{(f_i \circ \gamma_i)\sqrt{\dot{\gamma}_i}}$. Although this term has some stability to changes in $\gamma_i$, it does not achieve the desired invariance. Another approach is using a phase-invariant shape metric $d_s$ in a nonparametric model, see \cite{delicado2024comments}.

\end{itemize}
    

We modify past regression models using the Fisher-Rao Riemannian metric (FRM), termed $d_{FR}$, to create a new ScoSh model. $d_{FR}$ is phase invariant in the sense that $d_{FR}(f_1, f_2) = d_{FR}(f_1 \circ \gamma, f_2 \circ \gamma)$ for all warpings $\gamma$. The use of Square-Root Velocity Function (SRVF), specified later, simplifies the computation of $d_{FR}$. Under SRVF, the Fisher-Rao inner product becomes the $\ltwo$ inner product,  $\inner{f_1}{f_2}_{FR} = \inner{q_1}{q_2}_{\ltwo}$, where $q_i$s are the SRVFs of $f_i$s. This motivates an alternative term as a phase invariant inner product for the model. FRM’s invariance complicates parameter estimation, as the phases are nuisance variables that need to be removed through optimization, affecting parameter estimation.
Table~\ref{tab:list-models} lists a summary of the regression models and their acronyms used in this paper. The main contributions of this paper are: 
\begin{itemize}
\item It develops a new scalar-on-shape (ScoSh) regression model that uses the Fisher-Rao Riemannian metric to achieve invariance to the phase component of predictor functions. It solves the function registration (phase separation) inside the regression model rather than treating it as a preprocessing step. 

\item It introduces a concept of {\it regression phase} and {\it regression mean} associated with functional data. While the past definitions of mean shape and phase of in FDA (\cite{marron-etal-EJS:2014,marron-etal-Statsc:2015, SK2016}) are based on optimal alignments of peaks and valleys, the regression phase and mean result from those optimal time warpings that help minimize prediction error of the response variable. 

\item It uses classical index models (single and multiple) for enveloping the Fisher-Rao inner products to introduce nonlinear relationships in the model. 

\item It performs exhaustive experimental evaluations of the proposed model using simulated data (with known ground truths) and real data with interpretable solutions. The modeling performances compete successfully with state-of-the-art methods. 

\item The ScoF models can differ depending on the inner product between $\beta$ and $f_i$: the $\ltwo$ and Fisher-Rao inner products. The $\ltwo$ version is the commonly used FLM model, but we also include the Fisher-Rao version in the experiments for comparisons.
\end{itemize}

\section{Proposed Method}

The proposed scalar-on-shape (ScoSh) regression model requires the notion of shape in precise mathematical terms. First, we summarize the concept of \textit{shapes} of scalar functions and their treatments. We then introduce the proposed ScoSh model and its properties. In the process, we also provide a novel concept of \textit{Regression mean and phase}. We follow up with model estimation and a Bootstrap analysis of this estimator. 

\subsection{Background: Quantifying Shapes of Scalar Functions}
Let $\mathcal{AC}$  be the set of all absolutely-continuous functions on $[0,1]$ and  $\mathcal{AC}_0$ be a subset of ${\cal AC}$ that satisfies $f(0) = 0$. Also, let $\Gamma$ be the space of all boundary-preserving positive diffeomorphisms of the unit interval $[0,1]$ to itself, {\it i.e.}, $\Gamma:=\{\gamma:[0,1]\to[0,1] | \gamma(0)=0, \gamma(1)=1,\gamma \text{ is a diffeomorphism}\}$. $\Gamma$ forms the time-warping group, and the action of $\Gamma$ on  $\mathcal{AC}_0$ is the mapping $\mathcal{AC}_0 \times \Gamma \to \mathcal{AC}_0$ given by $(f, \gamma) \triangleq f \circ \gamma$.
The mapping $f \mapsto f \circ \gamma$ simply changes the {\bf phase} of $f$ but not its shape. Since the shape of $f$ is deemed unchanged by the mapping $f \mapsto f \circ \gamma$, we define $f \sim g$ to be an equivalence relation on $\mathcal{AC}_0$, where $g = (f \circ \gamma)$ for some $\gamma \in \Gamma$.  An equivalence class under this relation is given by: 
$[f] = \{ f \circ \gamma \vert \gamma \in \Gamma\}$.
Such an equivalence class uniquely represents a shape, and the set of all shapes is the quotient space ${\cal S} = \mathcal{AC}_0/\Gamma = \{ [f] \vert f \in \mathcal{AC}_0\}$. 


To develop a regression model similar to Eqn \ref{eq:classic-FLM} using elements of the shape space ${\cal S}$, we need an inner product on ${\cal S}$. As discussed in \cite{SK2016}, the classical $\ltwo$ inner product is unsuitable for shape analysis. Instead, we use the Fisher-Rao Riemannian metric with the required invariance properties. This metric is complex, and one uses the Square-Root-Velocity-Function (SRVF) representation (\cite{SK2016}) for simplification.
The SRVF of a function $f \in \mathcal{AC}$ is defined to be $q = Q(f) \triangleq \mbox{sign}(\dot{f}(t)) \sqrt{|\dot{f}(t)|}$. 
The mapping $Q: f \mapsto q$ is a bijection between $\mathcal{AC}_0$ and $\ltwo$, with the inverse map given by $Q^{-1}(q)(t) \triangleq f(t) = \int_0^t q(s) |q(s)|~ds$. Thus, the mapping $f \mapsto (f(0), q)$ is a bijection between the larger set $\mathcal{AC}$ and $\real \times \ltwo$. 

For any $f \in \mathcal{AC}_0$ and $\gamma \in \Gamma$, the SRVF of the composition $f \circ \gamma$ is given by $Q(f \circ \gamma) =(q \circ \gamma) \sqrt{\dot{\gamma}}$; We will denote it by $q \star \gamma$. For a shape class $[f] \subset \mathcal{AC}$, the corresponding subset in $\ltwo$ given by: $[q] = \{(q \star \gamma) | \gamma \in \Gamma\}$. 
There are several advantages to using SRVFs in shape analysis of functions. One is that the Fisher-Rao inner product between any two functions $f_1$ and $f_2$ is the $\ltwo$ inner product between their SRVFs, {\it i.e.}, $\inner{f_1}{f_2}_{FR} = \inner{q_1}{q_2}_{\ltwo}$ and the Fisher-Rao distance is $d_{FR}(f_1, f_2) = \| q_1 - q_2\|_{\ltwo}$, where $q_1, q_2$ are the SRVFs of $f_1, f_2$, respectively. (From hereon, we will use $\inner{\cdot}{\cdot}$ and $\| \cdot \|$ to denote the $\ltwo$ inner product and norm.) With this identification, the invariance property of the Fisher-Rao metric can be stated as: 
\begin{equation} \label{eq:SRVF-invariance}
\inner{q_1}{q_2} = \inner{q_1 \star \gamma}{q_2 \star \gamma},\ \mbox{or}\ \  
\|q_1 - q_2\| = \| (q_1 \star \gamma) - (q_2 \star \gamma)\|\ .
\end{equation}
This invariance property leads to a well-defined {\it shape metric} $d_S(q_1,q_2) \equiv \inf \limits_{\gamma \in \Gamma} \|q_1 - (q_2 \star \gamma)\|$. Expanding the square of $d_s$, we get $\inf\limits_{\gamma \in \Gamma} \left(\|q_1\|^2+\|q_2\|^2 -2\langle q_1,q_2 \star \gamma\rangle\right)=\|q_1\|^2+\|q_2\|^2-\sup\limits_{\gamma\in\Gamma}2\langle q_1,q_2\star\gamma\rangle$. This shows that if the norms of $q_1$, $q_2$ are constant, then $d_S$ is negatively proportional to the quantity: 
$\sup\limits_{\gamma\in\Gamma}\langle q_1,q_2\star\gamma\rangle$. This last term motivates the phase-invariant inner product in the proposed model.

\subsection{Proposed Scalar-on-Shape (ScoSh) Regression Model}

To focus on shapes of $\{f_i\}$, we need invariance to the phase of $\{f_i\}$, {\it i.e.}, replacing any $f_i$ with $f_i \circ \gamma_i$ should not change the response $y_i$. To achieve this, we use $\sup_{\gamma \in \Gamma}  \inner{\beta}{q_i \star \gamma}$ as a surrogate for $\inner{\beta}{f_i}$ in Eqn.~\ref{eq:classic-FLM}. The invariance of the Fisher-Rao inner product and the group structure of $\Gamma$ results in the property:
$\sup_{\gamma \in \Gamma}  \inner{\beta}{q_i \star \gamma} = 
\sup_{\gamma \in \Gamma}  \inner{\beta}{(q_i \star \gamma_0) \star \gamma}$,
for any $\gamma_0 \in \Gamma$. Thus, this expression is truly invariant to the phase of $f_i$ and depends only on its shape. To add flexibility to the model, we introduce two functions: (1) an index function $h: \real \to \real$, and (2) an offset function $g: \real \to \real$. We will assume that $h, g \in {\cal C}(\real, \real)$. The overall model can now be stated as: 
\begin{equation}\label{eq:Model}
y_i = g(f_i(0)) + h\left(\sup_{\gamma_i\in\Gamma}{\left\langle\beta,q_i\star\gamma_i\right\rangle}\right)+\epsilon_i\ ,\  i=1,\dots, n.
\end{equation}
Here $q_i, \beta \in \ltwo$, $\gamma_i \in \Gamma$, $g, h \in {\cal C}$, and $\epsilon_i \in \real$ are
{\it i.i.d.} from $N(0, \sigma^2)$. 
We will call this the Single-Index Scalar-on-Shape (SI-ScoSh) model. The parameters of this model are $\{\beta, h, g, \sigma^2\} \in \ltwo \times {\cal C} \times {\cal C} \times \real_+$. As a special case, we will also study when $h(x) = x$ and will call it the ScoSh model (without the SI prefix). 
Next, we discuss important properties of this model and impose conditions on the parameters to enforce identifiability. 

\begin{enumerate}

\item {\bf Fisher-Rao vs. $\ltwo$ Inner Product}: One might ask why not use $\sup_{\gamma_i \in \Gamma}  \inner{\beta}{f_i \circ \gamma_i}$ instead of $\sup_{\gamma_i \in \Gamma}  \inner{\beta}{q_i \star \gamma_i}$ in the model? The reason is that the former is degenerate and loses information about $f_i$. Mathematically, the issue is $\inner{f_1}{f_2} \neq \inner{f_1 \circ \gamma}{f_2 \circ \gamma}$. In contrast, the invariance property of SRVFs in Eqn.~\ref{eq:SRVF-invariance} is essential for this model.

\item {\bf Properties of the Supremum Term}: The term $\sup_{\gamma \in \Gamma}  \inner{\beta}{q_i \star \gamma}$ is not linear in $q_i$, due to the presence of the $\sup$ operation. Also, this term is non-negative, which limits its direct use in the regression model. However, using the index function $h$ allows for negative values of $y_i$s. 

\item {\bf Identifiability of $\beta$}:
Note that $\beta$ is defined only up to its equivalence class $[\beta]$ since,   
$\sup_{\gamma_i\in\Gamma} \inner{\beta}{q_i\star\gamma_i} = \sup_{\gamma_i\in\Gamma} \inner{\beta \star \gamma_0}{q_i\star\gamma_i}$, for any $\gamma_0 \in \Gamma$. To ensure uniqueness, we restrict ourselves to a specific element of this class, as follows: 
We impose an additional \textit{centering} condition on $\beta$  through the phases $\{ \widehat{\gamma}_i\}$. We require that $\frac{1}{n}\sum_{i=1}^n \widehat{\gamma}_i = \gamma_{id}$ (note that $\gamma_{id}(t) = t$), where $\widehat{\gamma}_i = \argmax_{\gamma_i \in \Gamma} \inner{\beta}{q_i\star\gamma_i}$.
Once all the $\widehat{\gamma}_i$s are computed, we can simply use their average $\bar{\gamma} = \frac{1}{n} \sum_{i=1}^n \widehat{\gamma}_i$ to center any estimate of $\beta$. 
In a standard FLM model (Eqn.~\ref{eq:classic-FLM}), the search for $\beta$ can be restricted to the span of $\{f_i\}$ since any component of $\beta$ lying in the orthogonal of the span is lost after the inner product. This simplification does not hold in the proposed model. Even when $h(x) = x$,  $\beta$ is an element of a much larger space: 
$
\mbox{span}\{ [q_i], i=1,\dots, n\} = 
\left\{ \sum_{i=1}^n a_i (q_i \star \gamma_i)~\vert~ \gamma_1, \dots, \gamma_n \in \Gamma,\  a_i\in\real \right\}$.

\item {\bf Identifiability of $h$}:
    Another degree of freedom is associated with the scale of the argument of $h$. Since 
    $h(\sup_{\gamma_i\in\Gamma}{\inner{\beta}{q_i\star\gamma_i}}) = 
    h(\frac{1}{a}\sup_{\gamma_i\in\Gamma}{\inner{a \beta}{q_i\star\gamma_i}})$,
for any $a \in \real_+$, this adds an ambiguity to the definition.  One can remove it by imposing a constraint such as $\int h(t)~dt = 1$, or if using a polynomial form, fixing a coefficient of $h$.


\item {\bf Identifiability of $g$}: 
We can resolve any ambiguity in $g$ by setting $g(0)=0$. 
\end{enumerate}
With these constraints, the model is fully specified, and the parameters are well-defined.

\subsection{Model Parameter Estimation}

Next, we study the problem of estimating model parameters from the observed data $\{ (f_i, y_i) \in {\cal AC} \times \real,~i=1,2,\dots, n\}$. We pre-compute the SRVFs $\{q_i\} \in \mathbb{L}^2$ of the predictor functions $\{f_i\}$.
Then, given the observations $\{(y_i, q_i, f_i(0)) \in \real \times \ltwo, i=1,\dots, n\}$, the inference problem is to estimate the quantities $h$, $g,\beta$, and $\sigma^2$ from the data.  To simplify estimation, we will express $\beta \in \ltwo$ using a truncated orthogonal basis ${\cal B} = \{b_j, j=1,\dots, J\}$ according to: 
$\beta(t) = \sum_{j=1}^J c_j b_j(t)$. ${\cal B}$ can be either a predefined basis, {\it e.g.}, the Fourier basis, or can be extracted from the training dataset through functional PCA. Then, the maximum-likelihood estimates of $h,g$ and ${\bf c} = \{c_j\}$ are given by: 
\begin{eqnarray}\label{eq:mle}
            (\widehat{\bf c}, \widehat{h},\widehat{g})&=& \argmin_{{\bf c} \in \real^J, h \in {\cal C}(\real,\real),g \in {\cal C}(\real,\real)} 
            H({\bf c}, g, h),\ \ \mbox{where}\ \\
            H({\bf c}, g, h) &\triangleq&
            \left[ \sum_{i=1}^n{\left\{y_i-g(f_i(0))-h\left(\sup_{\gamma_i\in\Gamma}\left\langle\sum_{j=1}^J{c_jb_j,(q_i\star\gamma_i)}\right\rangle\right) \right \}}^2 \right]\ . \nonumber
\end{eqnarray}
One can impose a roughness penalty on $\beta$ to control its smoothness, if needed.

        

\noindent {\bf Iterative Parmeter Estimation}: 
 To minimize $H$ with respect to $g$, $h$, and $\beta$, we use a coordinate-descent approach, optimizing one parameter at a time while fixing the others. Estimating $\sigma^2$ from the residual variance is straightforward and not discussed. Algorithm \ref{ghtoget} summarizes these steps with finer details about the estimation process presented in the \textcolor{blue}{Supplementary Material}.

\begin{breakablealgorithm}
    \caption{Estimation of $\beta$ keeping $h$ and $g$ 
    fixed}\label{algo:est-beta}
    \begin{algorithmic}[1]
        \State \textbf{Input} $\widehat{h},\widehat{g}.
        $, matrix of SRVF's $q=\{\tilde{q_1},\cdots,\tilde{q_n}\}$, basis functions $b_1(t),\cdots,b_J(t)$.
        \State \textbf{Initialize} ${\bf c}\in\mathbb{R}^J$. Compute initial $\widehat\beta(t)=\sum\limits_{j=1}^J c_j$.
        \State For each observation $i$, \textbf{ find} the optimum time warping function :  $\gamma_i'=arg\sup\limits_{\gamma_i\in\Gamma}\inner{\widehat\beta}{q_i\star\gamma_i}$ using the Dynamic Programming algorithm(~\cite{SK2016}).
        \State \textbf{Update} the SRVF's registering them to $\widehat\beta$ : $q_i'=q_i\star\gamma_i'$.
        \State Using an optimization method, (such as {\tt fminunc} or {\tt simulannealbnd} in MATLAB) \textbf{minimize} the cost function (\ref{eq:mle}): $\widehat{\bf c}=arg\min\limits_{c\in\mathbb{R}^J}{H({\bf c},\widehat{h},\widehat{g})}$.
        \State \textbf{Update} $\widehat{\beta}(t)=\sum\limits_j{\widehat{c}_j\cdot b_j(t)}$ and $q_i=q_i^\prime\ \forall i$.
        \State  \textbf{If}  $H\left(\widehat{\bf c},\widehat{h},\widehat{g}\right)$ is large, return to step 3, else go to step 8.
        \State To remove the extra degree of freedom in $\beta$,\textbf{ compute} $\bar{\gamma}=\frac{1}{n}\sum\limits_i \gamma'_i$.
        \State Obtain the estimate $\widehat\beta=\widehat\beta\circ\bar{\gamma}^{-1}$.
    \end{algorithmic}
\end{breakablealgorithm}

\begin{breakablealgorithm}
    \caption{Elastic shape regression model} \label{ghtoget}
    \begin{algorithmic}[1]
        \State \textbf{Initialise} $\widehat{h}(x)=h_0(x)$ , $\widehat{g}(x)=0$.
        \State Given $\widehat{h},\widehat{g}$, \textbf{estimate} $\widehat\beta$ using Algorithm \ref{algo:est-beta}.
        \State Once obtained $\widehat\beta$, create $y_i'=y_i-\widehat{g}(f_i(0))$ and \textbf{estimate} $\widehat h$ using 
        \begin{itemize}
            \item \textbf{Define} estimated inner product as  $\widehat{y_i}=\sup\limits_{\gamma_i\in\Gamma}{\langle\widehat\beta,q_i\star\gamma_i\rangle}$.
            \item \textbf{Fit} a polynomial or a non-parametric curve $\widehat h$ between the responses $y_i'$'s and the estimated inner products $\widehat{y_i}$'s.
        \end{itemize}
        \State Remove the scaling degree of freedom from our estimate (by fixing the highest coefficient of $h$ to 1 and adjusting the other coefficients of $h$ and all of $\beta$ accordingly).
        \State  With $y''_i=y_i-\widehat{h}(\widehat{y}_i)$ \textbf{calculate} $\widehat g$ using 
        \begin{itemize}
            \item \textbf{Fit} a quadratic polynomial $\widehat{g}$ on the $y''$s. (As explained in Appendix 6.1, we restrict our search for optimal $g$ to a quadratic polynomial).
        \end{itemize}
        \State Iterate steps 3 to 5 until H converges.
        \State If $H\left(\widehat {c},\widehat h,\widehat{g}\right)$ is small, then stop;  else return to step 2.
    \end{algorithmic}
\end{breakablealgorithm}

\subsection{Estimator Analysis Using Bootstrap Sampling}

The estimators of $\beta$, $h$, and $g$ haev been defined using a joint optimization problem (Eqn.~\ref{eq:mle}) involving multiple parameters and nuisance variables. Ideally, one would like the distributions of estimated quantities for bias and consistency analysis. Several asymptotic distributions of $\beta$ and $h$ have been derived for FLM and related models ({\it e.g.}, \cite{YL2010}, \cite{JM2015}). However, estimating regression parameters in the shape context is much more difficult. The cost function, which includes a supremum over the nuisance variables $\{\gamma_i\}$, is nonlinear and complex. $\Gamma$ is an infinite-dimensional, nonlinear manifold, adding to the complexity. Additionally, Eqn.~\ref{eq:Model} has a potentially nonlinear index function $h$, complicating prediction error analysis.
\cite{du2015size} developed a theory for regression modeling and analysis in shape matching, but their context differs from our functional data setting.

Lacking analytical distributions, we take a computational approach and rely on bootstrap sampling. Bootstrapping allows us to examine estimator properties ({\it e.g.}, variance) by sampling with replacement and approximating the distribution of estimators ($\widehat{\beta}, \widehat{h}, \widehat{g}$). We will empirically analyze these estimators by generating numerous bootstrap replicates.

    \begin{figure}[!ht]
        \centering
        
        \begin{tabular}{ccc}
            \includegraphics[width=5cm]{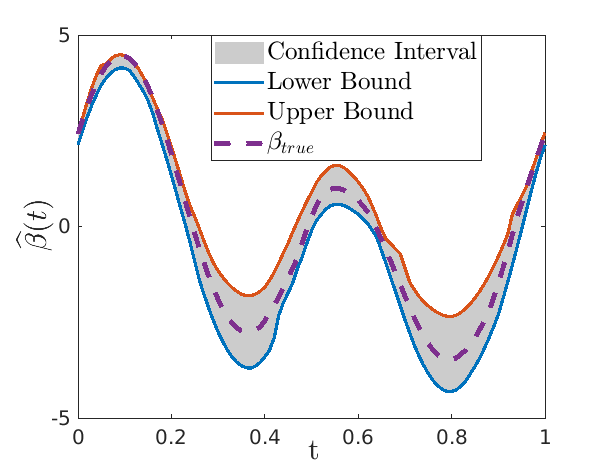} &
            \includegraphics[width=5cm]{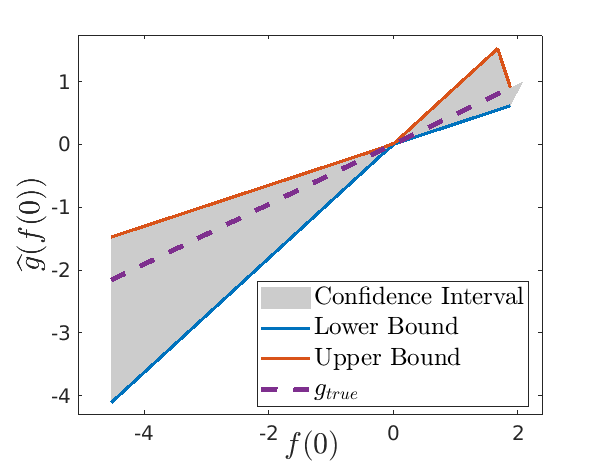} &
            \includegraphics[width=5cm]{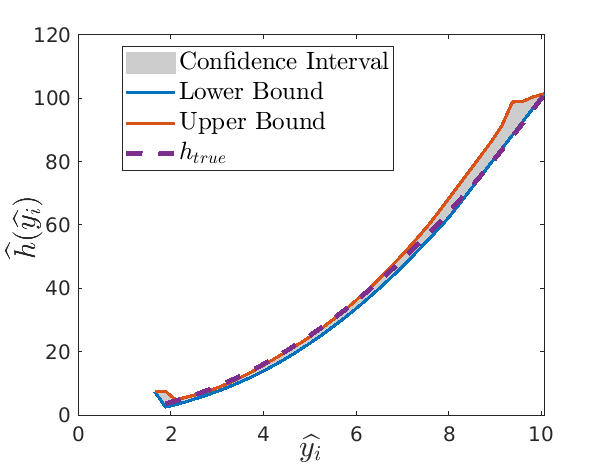} \\
        \end{tabular}
        \begin{tabular}{cc}
            \includegraphics[width=6cm]{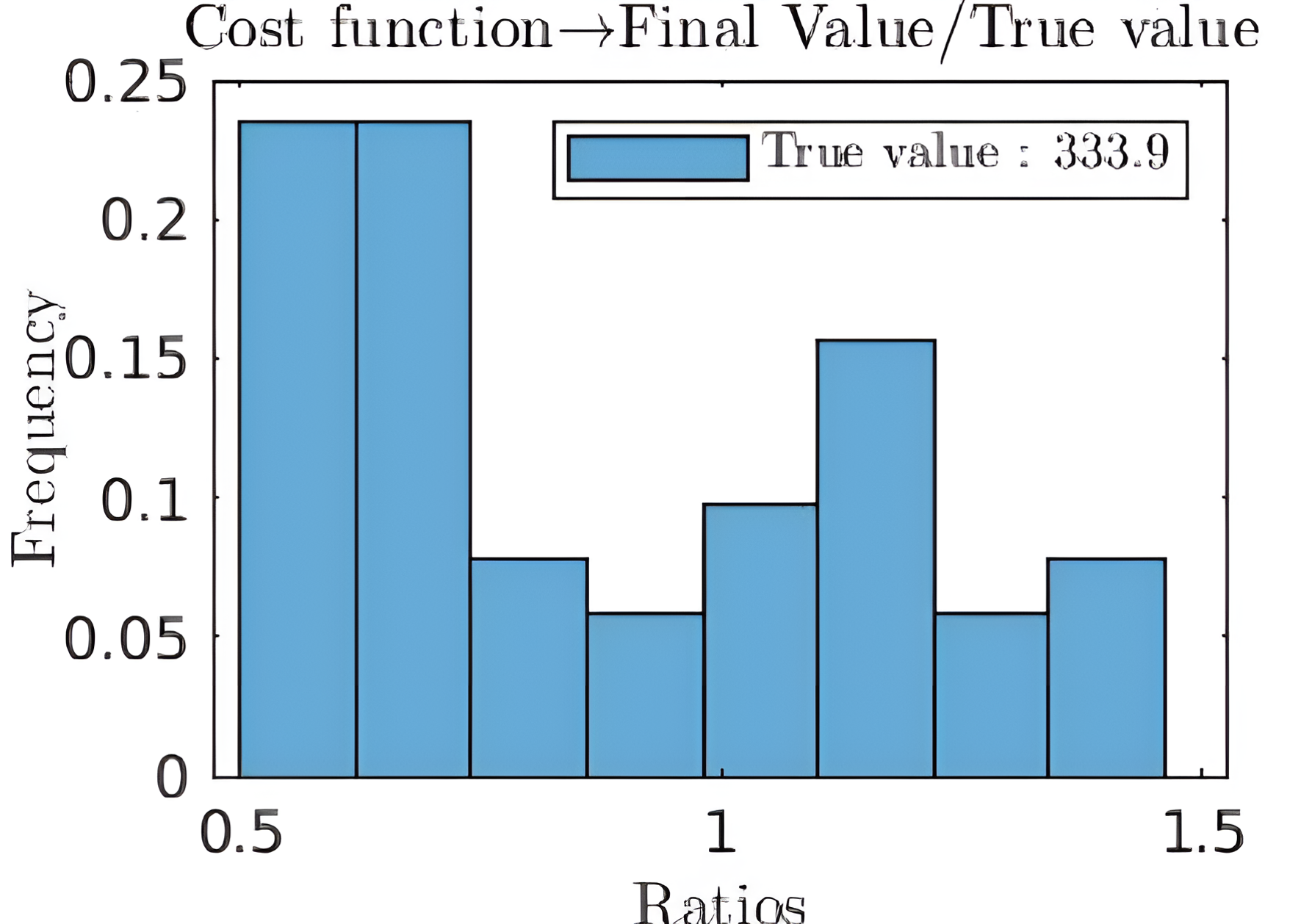} &
            \includegraphics[width=6cm]{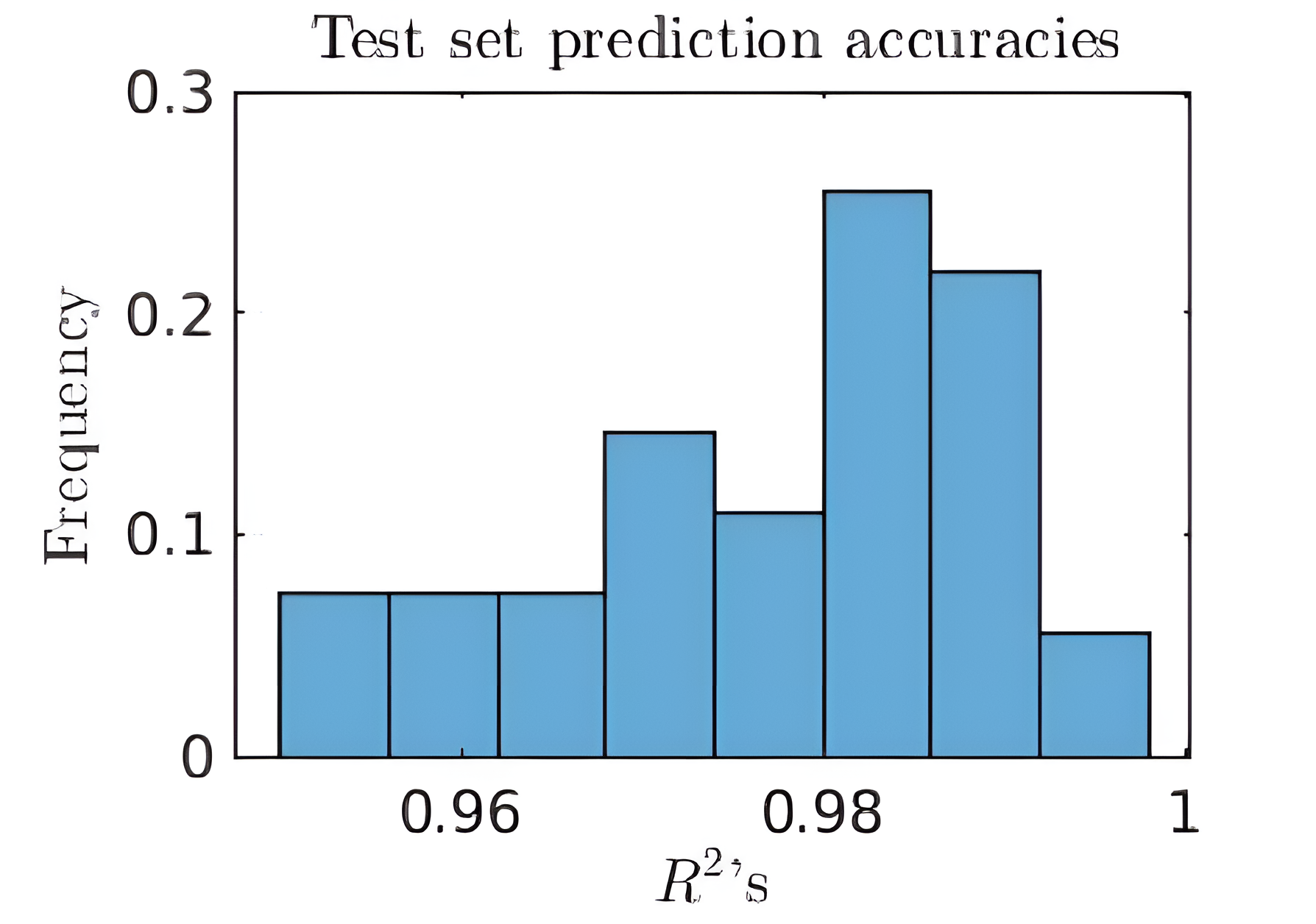}
        \end{tabular}
        \caption{Top: The three panels plot  $95\%$ bootstrap confidence intervals of the estimated quantities $\widehat\beta,~\widehat g$, and $\widehat h$, and their ground truth values. Bottom: The left panel shows a histogram of the ratio of $H_{final}$ and $H_{true}$ for bootstrap samples, and the right panel plots a histogram of the $R^2$ values of these samples.} 
        \label{CI}

    \end{figure}

To illustrate this approach, we conducted an experiment with parameters: $h(x)=x^2-x$, $g(x)=x$, and $\beta$ as shown in Fig.~\ref{CI} top-left, and data simulated from Eqn.~\ref{eq:Model} (simulation details are 
provided later in Section 3). To evaluate estimator performance using Bootstrap, we generated 100 randomizations of train-test sets, performed estimation using Algorithm 2, and evaluated performance. From the bootstrap replicates, we computed 95\% confidence intervals and compared them to the true values.
Fig.~\ref{CI} shows $\widehat{\beta}$, $\widehat{g}$, and $\widehat{h}$ from left to right. The gray regions depict the 95\% confidence intervals, with red and blue curves denoting the bounds and dotted curves representing true values. These plots show that the true values of $\beta$, $h$, and $g$ lie within the confidence intervals, validating our numerical approach.

The bottom-left shows a histogram (from 100 bootstrap samples) of the ratio $\frac{H_{final}}{H_{true}}$, where $H_{final}$ is the converged value of $H$ and $H_{true}$ is the value of $H$ for ground truth parameters. We can see that the final $H$ values converge to within $0.5 - 1.5$ times the true value of the cost function. This underscores the good convergence properties of our gradient approach. The bottom-right histogram shows $R^2$ values (prediction accuracy) on test data for each of the 100 model fits, highlighting the excellent prediction performance of the estimated model.

\subsection{Regression Phase and Regression Mean}

Our estimation of model parameters involves aligning predictor SRVFs $\{q_i\}$ to the coefficient $\beta$ using time warpings $ \gamma_i$ during estimation. This perspective allows us to define the phase components of $f_i$s in a different way than the traditional phase-amplitude separation. 

\begin{figure}[t]
\begin{center}
    \begin{tabular}{|c|c|c|}
    \hline
    \includegraphics[height=1.9in]{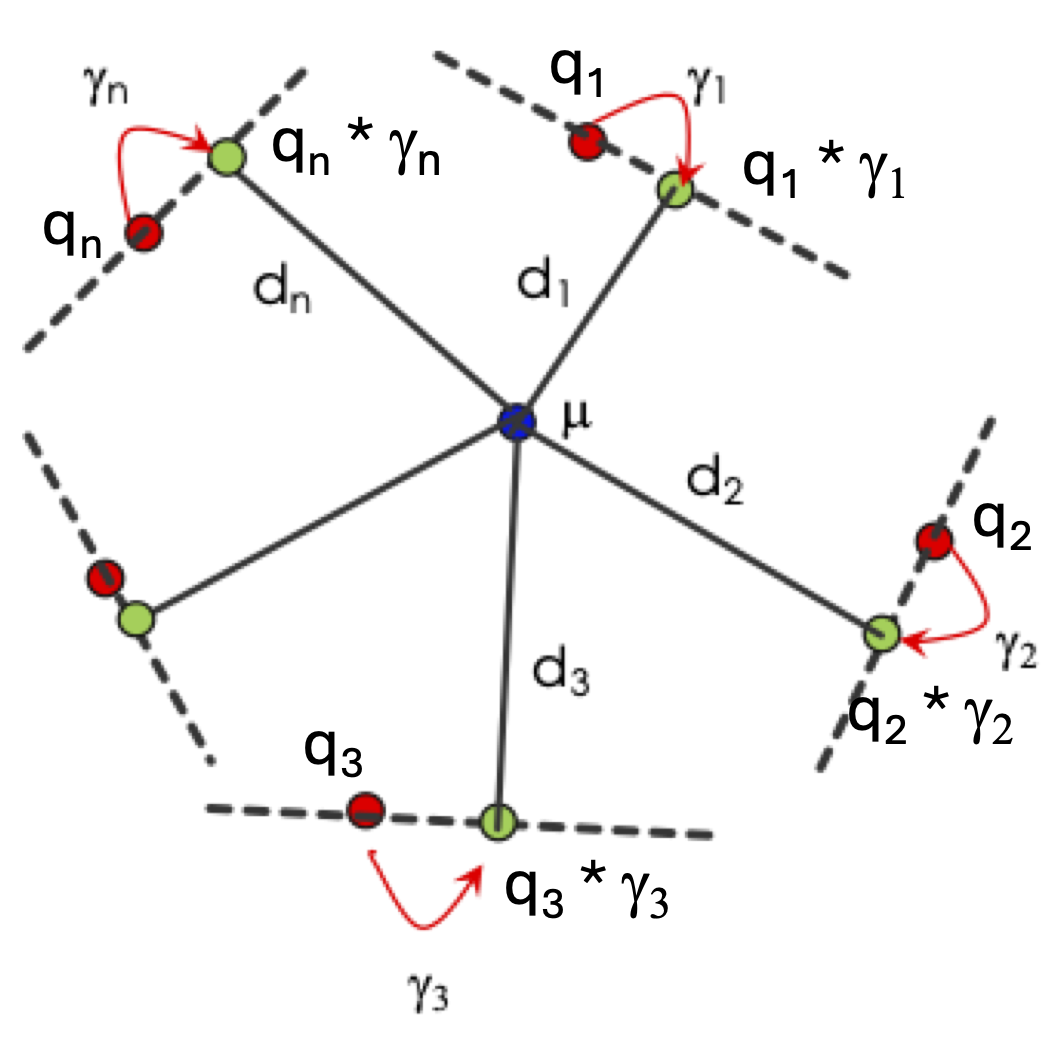}&
    \includegraphics[height=1.9in]{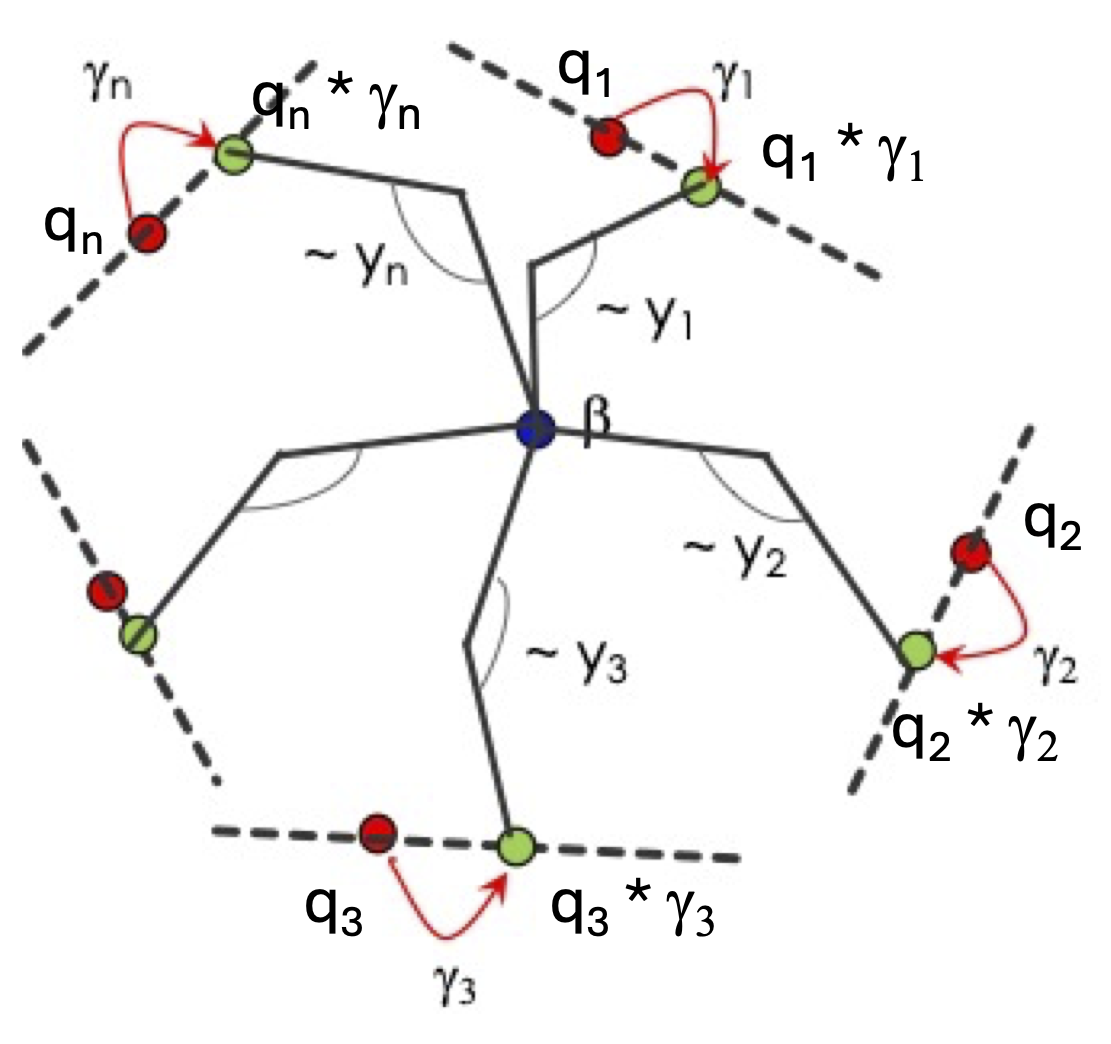} &
    \raisebox{0.23in}{\includegraphics[height=1.6in]{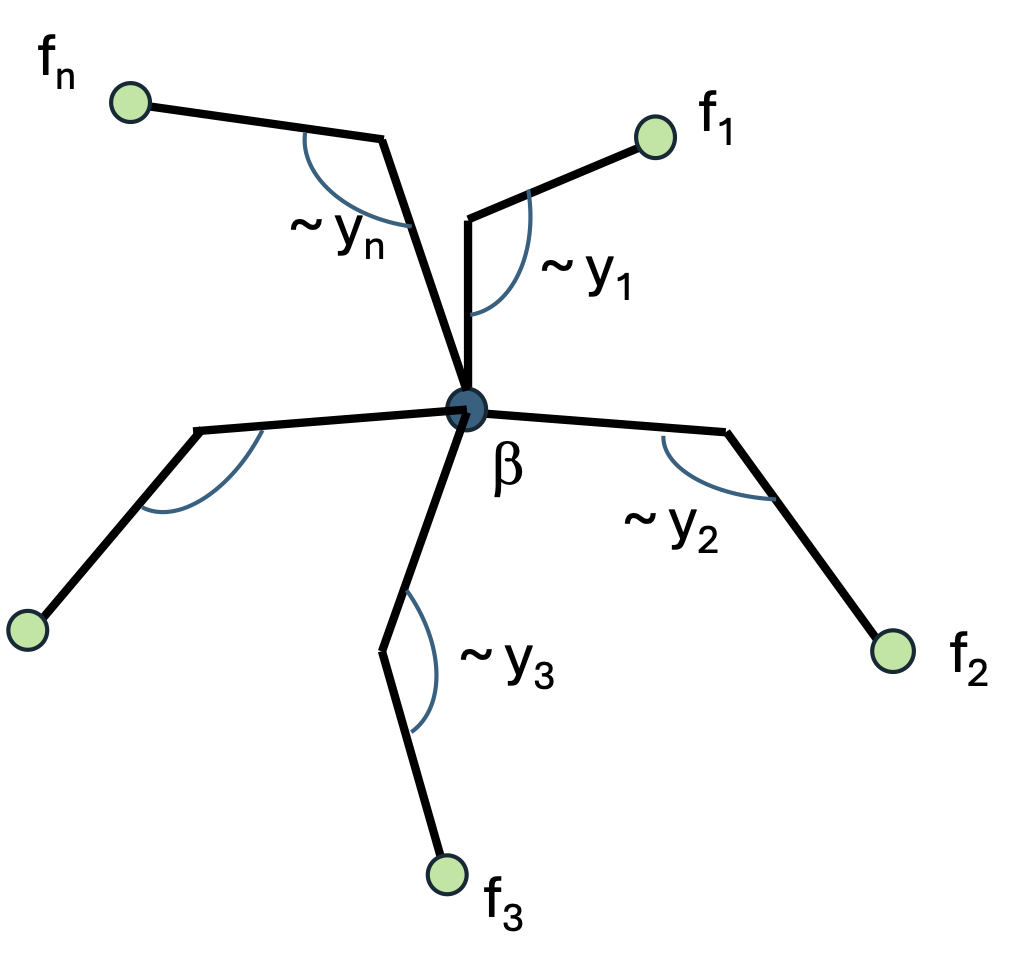}}\\
    \hline
    $\gamma_i = \arginf \| \mu - (q_i \star \gamma)\|^2$ & 
    $y_i \approx \sup_{\gamma_i} \inner{\beta}{q_i \star \gamma_i}$ &
    $y_i \approx \inner{\beta}{f_i}$\\
    \hline
    Shape Registration & 
    Registration and Regression (ScoSh) & Regression (ScoF) \\
    \hline
    \end{tabular}
    \caption{Left: Alignment and shape averaging of functions using SRVFs; Middle:  Alignment of SRVFs to regression coefficient $\beta$ to approximate responses $y_i$ through inner products; Right: Approximation of responses $y_i$ through inner products without any alignment.} \label{fig:register}
\end{center}
\end{figure}

\noindent {\bf Classical Phase-Amplitude Separation}: In the past work (\cite{TUCKER201350,marron-etal-EJS:2014,marron-etal-Statsc:2015, SK2016,zhang-etal}), the \textit{phases} of functions have been defined as the time-warpings required to align their peaks and valleys.  Mathematically, the phase for a function $f_i$ (with SRVF $q_i$) is defined as $\widehat{\gamma}_i = \arg\sup_{\gamma\in\Gamma}\inner{\mu}{q_i\star\gamma}$, where $\mu \in \ltwo$ is the Karcher or the Fr\'{e}chet mean of the given functions and is defined using: 
\begin{eqnarray}
    \mu &=& \arg\inf_{q \in \ltwo} \sum_{i=1}^n \left( \inf_{\gamma_i \in \Gamma} \| q - q_i \star \gamma_i\|^2 \right) \nonumber \\
    &=& \arg\inf_{q \in \ltwo} \sum_{i=1}^n \left( \|q\|^2 + \|q_i\|^2 - 2 \sup_{\gamma_i \in \Gamma} \inner{q}{q_i \star \gamma_i} \right). \label{eqn:comp1}
\end{eqnarray}
Note that $\{\widehat{\gamma}_i\}$ are defined through the optimization in Eqn \ref{eqn:comp1}. The left panel of Fig.~\ref{fig:register} shows a cartoon example of this idea, where SRVFs $\{q_i\}$ are warped into $\{(q_i \star \gamma_i)\}$ to align with the current estimate of the shape average $\mu$. 

\noindent {\bf Regression-Based Function Aligment}: 
Similarly, we define optimal time-warping in the ScoSh model using $\widehat{\gamma}_i= arg\max_{\gamma\in\Gamma}\inner{\widehat{\beta}}{q_i\star\gamma}$, where the estimator of $\beta$ is: 
\begin{eqnarray}
    \widehat{\beta} &=& \arg\inf_{\beta \in \ltwo} \sum_{i=1}^n \left( y_i - g(f_i(0)) - h(\sup_{\gamma_i \in \Gamma} \inner{\beta}{q_i \star \gamma_i}) \right)^2 \nonumber \\
    &=& \arg\inf_{\beta \in \ltwo} \sum_{i=1}^n \left( y_i - \sup_{\gamma_i \in \Gamma} \inner{\beta}{q_i \star \gamma_i} \right)^2,\ \mbox{assuming } h(x)=x,~~g = 0\ .
    \label{eqn:comp2}
\end{eqnarray}
Comparing Eqns.~\ref{eqn:comp1} and \ref{eqn:comp2}, we see the parallels between $\mu$ and $\widehat{\beta}$.  In Eqn.~\ref{eqn:comp1}, one seeks a $\mu$ that is closest to all $q_i \star \widehat{\gamma}_i$, and in the process making $2 \sup_{\gamma_i \in \Gamma} \inner{q}{q_i \star \gamma_i}$ as close to $\|\mu\|^2  + \|q_i\|^2$ as possible. Similarly, in Eqn.~\ref{eqn:comp2}, the optimal $\widehat{\beta}$ makes $\sup_{\gamma_i \in \Gamma} \inner{\beta}{q_i \star \gamma_i}$ as close to $y_i$ as possible (assuming $h(x)= x$, $g = 0$). This motivates naming  $\widehat{\beta}$ as the {\it regression mean} of the shapes of $\{f_i\}$ {\it w.r.t} responses $\{y_i\}$. The middle panel of Fig.~\ref{fig:register} shows a cartoon illustration of this idea. The right depicts a ScoF or FLM model where one approximates responses $\{y_i\}$ using the inner products between $\{f_i\}$ and $\beta$ without any alignment.

        \begin{figure}[!ht]
            \hspace{-0.6cm}
            
            \begin{tabular}{|cccccc|}
            \hline
                \includegraphics[scale=0.3]{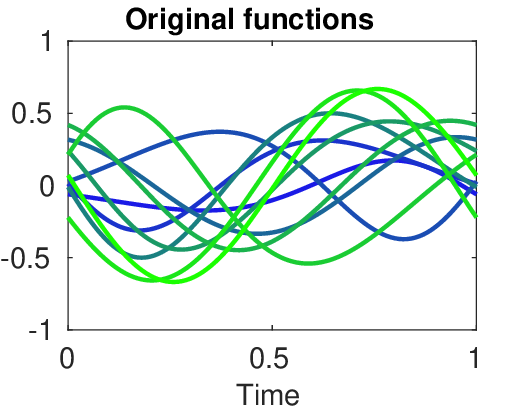}\hspace{-0.5cm} &
                \includegraphics[scale=0.3]{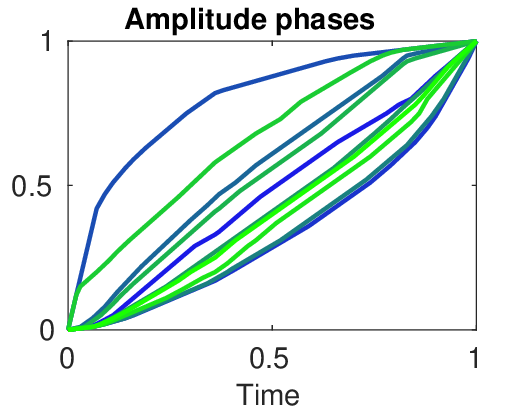}\hspace{-0.5cm} &
                \includegraphics[scale=0.3]{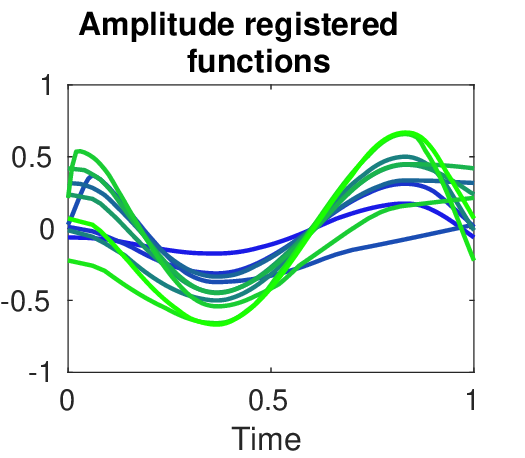}\hspace{-0.5cm} &
                \includegraphics[scale=0.3]{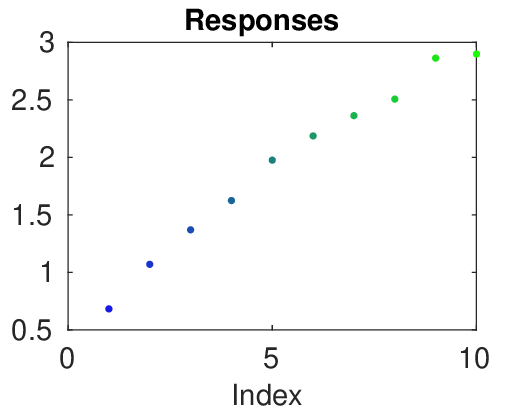}\hspace{-0.5cm} &
                \includegraphics[scale=0.3]{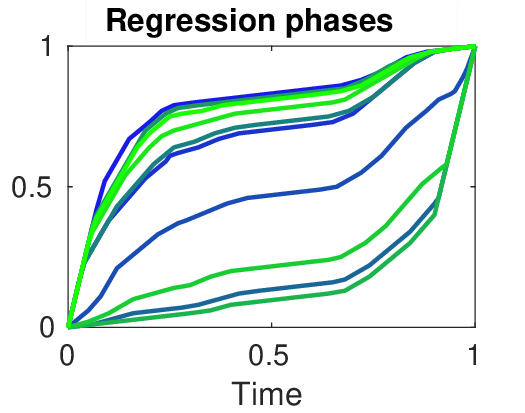}\hspace{-0.5cm} &
                \includegraphics[scale=0.3]{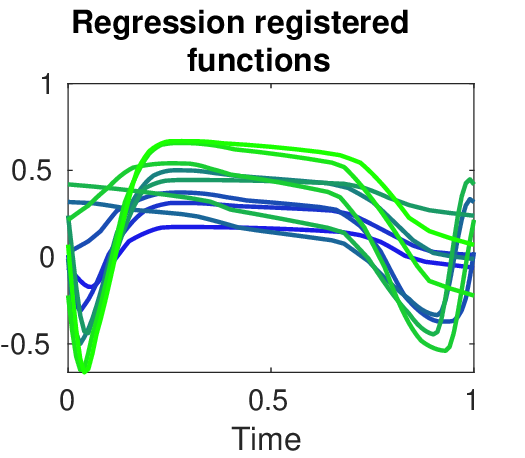}\\
                \hline
                \includegraphics[scale=0.3]{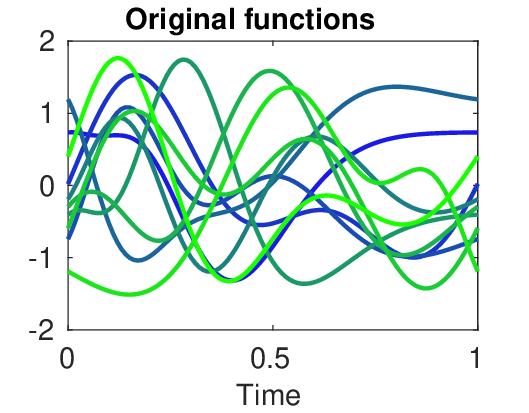}\hspace{-0.5cm} &
                \includegraphics[scale=0.3]{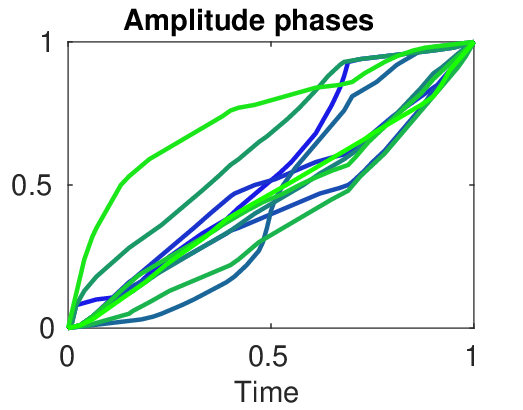}\hspace{-0.5cm} &
                \includegraphics[scale=0.3]{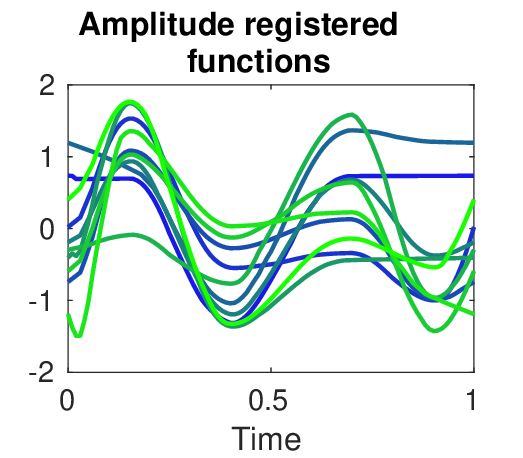}\hspace{-0.5cm} &
                \includegraphics[scale=0.3]{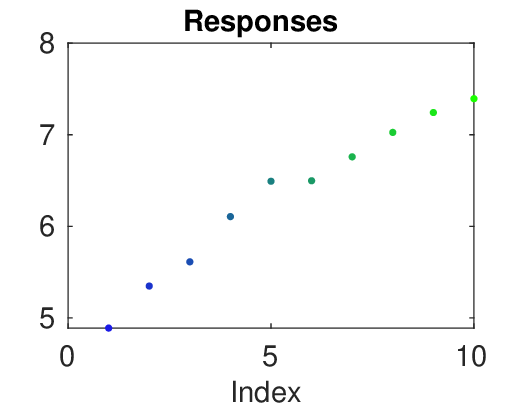}\hspace{-0.5cm} &
                \includegraphics[scale=0.3]{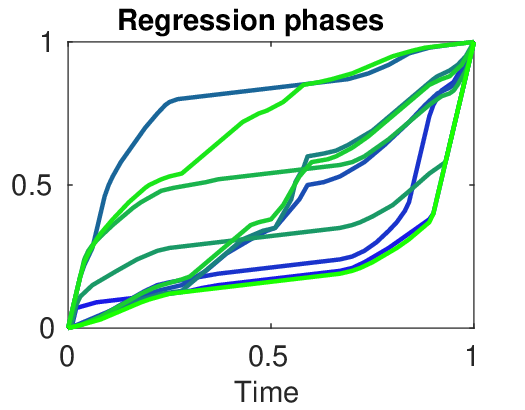}\hspace{-0.5cm} &
                \includegraphics[scale=0.3]{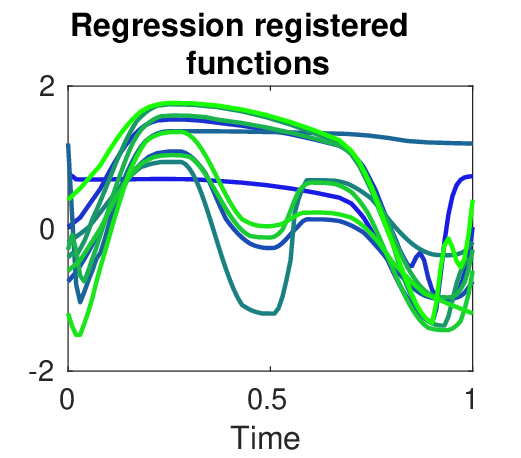}\\
                \hline
            \end{tabular}
            \caption{Two examples contrasting amplitude-phase and regression phase. In each row, the leftmost panel shows the original $\{f_i\}$, the second shows amplitude phases, and the third shows the aligned functions. The fourth panel shows response data $\{y_i\}$, the fifth shows regression phases, and the last shows regression registered functions.}
            \label{REGph}
        \end{figure}

\normalsize
Next, we present two simple examples in Fig.~\ref{REGph} to illustrate and compare regression means and amplitude means. Each row shows a different example. The simulation setup for these examples is the same as in Section \ref{sec:simulation}.  Here $f_i$'s are constructed using simple Fourier basis, $h$ and $g$ are both lower-order polynomials, and true $\beta$ is made of five Fourier bases. The traditional phase-amplitude separation seeks to align peaks and valleys in $\{f_i\}$, while the regression-based separation tries to match the inner product of $\beta$ and $(q_i \star \gamma_i)$ with $y_i$. The results naturally show significant differences in the phases of the two approaches.

\section{Experimental Results: Simulated Data} \label{sec:simulation}
        In this section, we simulate several datasets and use them to evaluate the proposed as well as some current models.

        \noindent {\bf Simulation Setup}:  In this experiment, we generate $f_i^0(t)=c_{i,1}\sqrt{2}\sin(2\pi t)+c_{i,2}\sqrt{2}\cos(2\pi t)$, where $c_{i,1},c_{i,2}\sim\mathcal{N}(0,1^2)$. To create predictors with arbitrary phases, we perturb each of these $f^0$'s by random $\gamma_i$'s : $f_i(t)=f_i^0 \circ \gamma_i$, where $\gamma_i(t)=t+\alpha \cdot t(T-t)\ \{t\in[0,T],\ \alpha\in U(-1,1)\}$.
        We calculate the corresponding SRVF's ($q_i$'s) of each of these $f_i$'s using $q_i = \mbox{sign}(\dot{f}_i(t)) \sqrt{|\dot{f}_i(t)|}$.  To define coefficient vector $\beta$, we use first $J$ elements of the Fourier basis $\{b_j\}=\{\sqrt{2} \cos(2\pi jx), \sqrt{2} \sin(2\pi jx), j=1,2\dots, J/2\}$ and some fixed coefficients $\tilde{c_0}=\{ 1,\cdots, 1\}$.  Also, we use low-order polynomials for $h$ (listed in the experiments) and a fixed $g(x)=x^2-1$. Then, we calculate responses $y_i$'s by adding $\epsilon_i \sim \mathcal{N}(0,0.01^2)$ as per Eqn.~\ref{eq:Model}.  For a sample size of $n=100$, we use 80\% of the dataset for training and the rest for testing in a five-fold cross-validation. For each random split, we use Algorithm~2  to estimate the model parameters. 

        \begin{figure}[!ht]
            \centering
            \begin{tabular}{cccc}
                \hspace{-1cm}\includegraphics[scale=0.5]{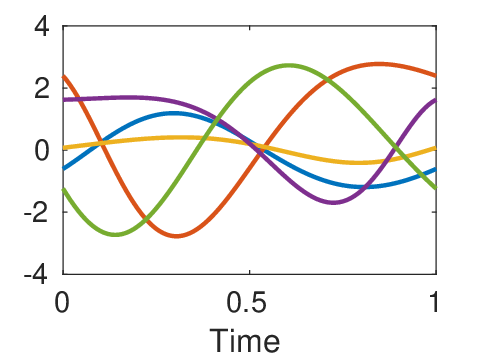}&
                \hspace{-0.5cm}\includegraphics[scale=0.5]{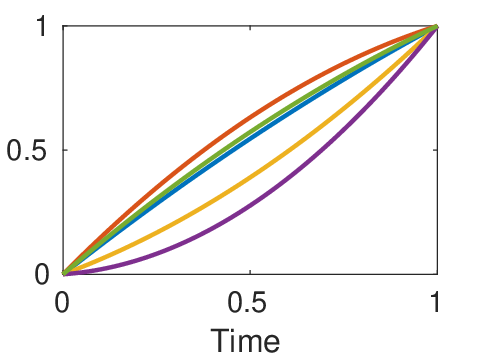}&
                \hspace{-0.5cm}\includegraphics[scale=0.5]{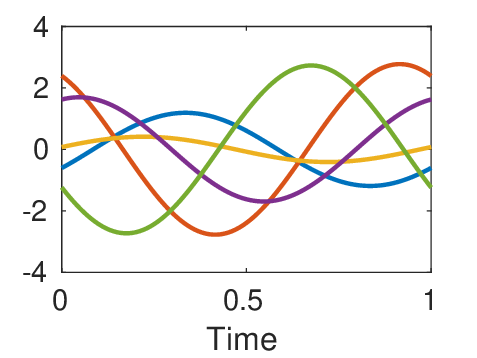}& 
                \hspace{-0.5cm}\includegraphics[scale=0.5]{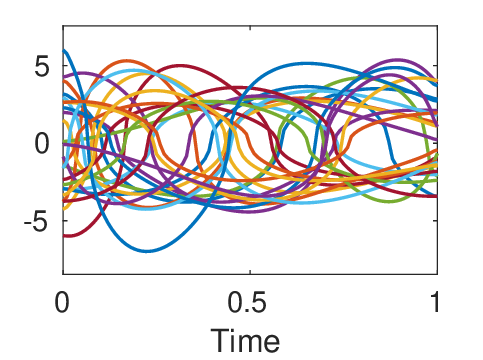}\\
                \hspace{-1cm}\includegraphics[scale=0.5]{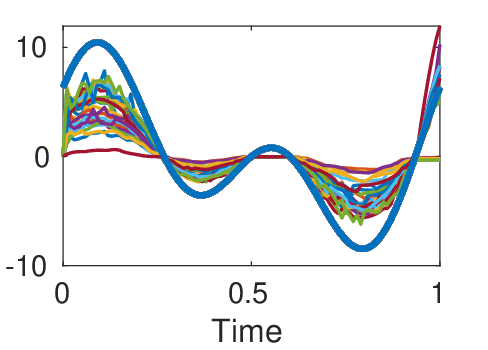}&
                \hspace{-0.5cm}\includegraphics[scale=0.45]{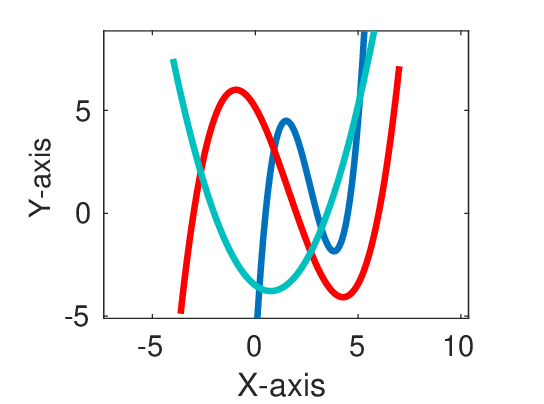}&
                \hspace{-0.5cm}\includegraphics[scale=0.45]{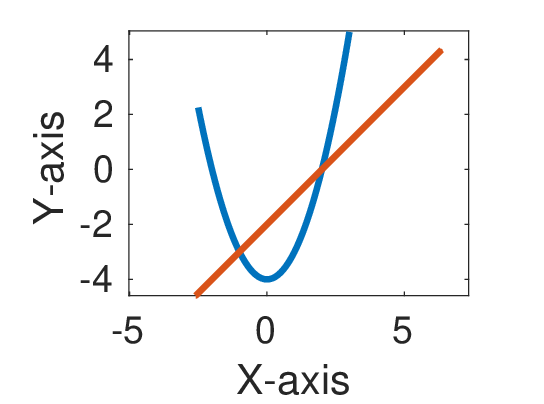}&
                \hspace{-0.5cm}\includegraphics[scale=0.5]{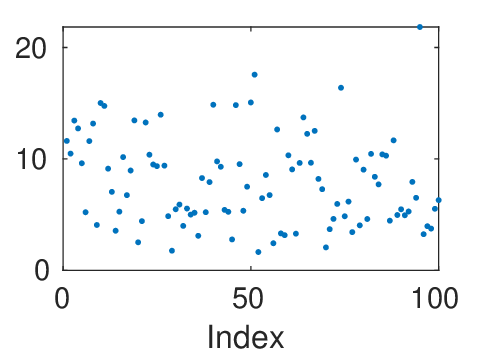}\\
            \end{tabular}
            \caption{Simulating experimental data. Top: From left to right, some initial predictor functions $\{f_i^0\}$ formed using a Fourier basis, random time-warpings $\{\gamma_i\}$, time-warped functions $\{f_i = f_i^0 \circ \gamma_i\}$, and their SRVFs $\{q_i\}$. Bottom: SRVF's after registering with a $\beta$ (blue line), some index functions $h$, some polynomial offset functions $g$, and the responses $\{y_i\}$ generated after adding random noise.} 
            \label{fig:enter-label}
        \end{figure}

        \noindent {\bf Model Comparisons}:
        Next, we compare performance of the ScoSh model with three other models (refer to Table~\ref{tab:list-models} for model acronyms and specifications): (1) SI-ScoF(\textit{FR}), which uses the functions without alignment, (2) ScoSh, which uses SRVFs with alignment but sets $h$ as identity, and (3) ScoF(\textit{FR}), resembling the classical FLM but using the Fisher-Rao inner product.
        During estimation, SI-ScoSh iteratively optimizes over $\beta$, $h$, and $g$ while registering functions. SI-ScoF(\textit{FR}) optimizes over $\beta$, $h$, and $g$ without registration. ScoSh includes registration and optimization over $\beta,\ g$. ScoF(\textit{FR}) estimates $\beta,\ g$.

        \subsection{Evaluating Response Prediction}
        We sequentially generate data from one of these stated models, apply all the models to that data, and quantify model performances using five-fold validation. The original model is naturally expected to perform the best, but comparing the performances of others is also informative. We quantify prediction performance using the $R^2  \left( = 1-\frac{\sum_i{(y_i-\widehat{y}_i)^2}}{\sum_i{(y_i-\bar{y})^2}}\right)$ statistic ($\bar{y}$ is the mean of the $y_i$'s and $\widehat{y}_i$ is the predicted value of $y_i$'s ). In the tables, columns represent different polynomial choices for true $h$ and numbers of basis functions (J) for true $\beta$, while rows correspond to different fitted models. The entries in cells are the means of $R^2$ values over five-fold replications, with standard deviations in parenthesis.  Additional tables can be found in the \textcolor{blue}{Supplementary Material}.

\noindent 1. {\bf Data from SI-ScoSh model}: 
        The left part of Table~\ref{SI-ScoSh} shows results for data generated from the SI-ScoSh model; this model has a nonlinear predictor-response relationship and non-informative predictor phases. The first two rows show SI-ScoSh results for different estimators of $h$, both providing very high $R^2$ values. ScoSh model, {\it i.e.},  $h(x) = x$, performs increasingly worse as the true $h$ becomes more complex. SI-ScoF(\textit{FR}) 
        captures some predictor-response relation but is inferior to ScoSh models. ScoF(\textit{FR}) performs much worse, indicating the need to remove nuisance phase variability for effective performance. Note that a negative $R^2$ means that predicted values are worse than the fixed guess $\bar{y}$.

        \begin{table}[!ht]
            \centering\small
            \caption{Test ($R^2)$ prediction performance comparison for data generated from SI-ScoSh (left three columns) and SI-ScoF(\textit{FR})(right three columns). linear: $h_{true}(x) = 3x-2$, quadratic: $h_{true}(x)= x^2-3x+2$, cubic: $h_{true}(x)= (x-0.5)(x-3)(x-4.5)$. $\beta_{true}[J=4]=\sum\limits_{i=1,3}2b_i+\sum\limits_{i=2,4}\sqrt{2}b_i$ and $\beta_{true}[J=6]=\sum\limits_{i=1,3,5}2b_i+\sum\limits_{i=2,4,6}\sqrt{2}b_i$.}
            \begin{tabular}{|c|||c|c|c||c|c|c|}
            \hline
                & \multicolumn{3}{|c||}{SI-ScoSh} & \multicolumn{3}{|c|}{SI-ScoF(\textit{FR})}\\
                \hline
                $h_{true}$ & linear & quadratic & cubic & linear & quadratic & cubic\\
                \hline
                $J(of \ \beta_{true})$ & 4 & 6 & 4 & 4 & 6 & 4\\
                \hline\hline
                 SI-ScoSh: Poly & 0.96(0.02) & 0.98(0.01) & 0.98(0.01) & 0.92(0.05) & 0.89(0.06) & 0.87(0.05)\\
                 \hline
                 SI-ScoSh: SVM & 0.97(0.01) & 0.98(0.01) & 0.97(0.01) & 0.94(0.02) & 0.90(0.04) & 0.89(0.04)\\
                \hline
                 SI-ScoF(\textit{FR})& 0.72(0.09) & 0.48(0.26) & 0.23(0.30) & 0.99(0.01) & 0.99(0.01) & 0.99(0.01)\\
                 \hline
                 ScoSh & 0.94(0.02) & 0.72(0.10) & 0.50(0.21) & $<0$ & $<0$ & $<0$\\
                 \hline
                 ScoF(\textit{FR})& $<0$ & $<0$ & $<0$ & $<0$ & $<0$ & $<0$ \\
                 \hline
            \end{tabular}            
            \label{SI-ScoSh} 
        \end{table}

\noindent 2. {\bf Data from SI-ScoF model}: 
        The right part of Table~\ref{SI-ScoSh} shows prediction performance for data from the SI-ScoF(\textit{FR}) model -- 
        a nonlinear index function $h$ and an informative phase component. As expected, SI-ScoF(\textit{FR}) performs best, with SI-ScoSh also doing well. ScoSh performs poorly, indicating the importance of the index function. Also, optimizing over $\{\gamma_i\}$ loses informative phase components and reduces performances. Prediction performance decreases from left to right as the complexity of $h$ increases.

 \noindent 3. {\bf Data from ScoSh model}: 
 Table~\ref{ScoSh} shows prediction performances for data from the ScoSh model, with $h(x) = x$ and non-informative phase components. Both ScoSh and SI-ScoSh give accurate predictions, with SI-ScoSh being a generalization of ScoSh. SI-ScoF(\textit{FR}), despite keeping nuisance phases, performs decently as the index function helps compensate for the mismatch. The ScoF(\textit{FR}) model, which keeps the nuisance phases but does not use an index function, performs poorly.
        \begin{table}[!h]
            \centering\small
            \begin{minipage}{0.45\textwidth}
                \centering\small
                \caption{Test ($R^2)$ prediction performance comparison for data generated from ScoSh.}
                \begin{tabular}{|c|||c|c|}
                \hline
                    $J(of \ \beta)$ & 4 & 6 \\
                    \hline\hline
                     SI-ScoSh: Poly & 0.98(0.01) & 0.99(0.01) \\
                     \hline
                     SI-ScoSh: SVM & 0.97(0.01) & 0.98(0.01) \\
                     \hline
                     SI-ScoF(\textit{FR}) & 0.83(0.04)& 0.68(0.2) \\
                     \hline
                     ScoSh & 0.98(0.01) & 0.96(0.03) \\
                     \hline
                     ScoF(\textit{FR}) & $<0$ & $<0$ \\
                     \hline
                \end{tabular} 
                \label{ScoSh}
            \end{minipage}\hspace{1cm}
            \begin{minipage}{0.45\textwidth}
            \centering\small
            \caption{Test ($R^2)$ prediction performance comparison for data generated from ScoF(\textit{FR})}
                \begin{tabular}{|c|||c|c|}
                \hline
                     $J(of \ \beta)$ &4 & 6 \\
                    \hline\hline
                     SI-ScoSh: Poly & 0.91(0.05) & 0.92(0.05) \\
                     \hline
                     SI-ScoSh: SVM & 0.94(0.03) & 0.92(0.03) \\
                     \hline
                     SI-ScoF(\textit{FR}) & 0.99(0.01) & 0.99(0.01) \\
                     \hline
                     ScoSh & $<0$ & $<0$ \\
                     \hline
                     ScoF(\textit{FR}) & 0.99(0.01) & 0.99(0.01) \\
                     \hline
                \end{tabular}
                 
                \label{ScoF}
            \end{minipage}    
        \end{table}    
    
   \noindent 4. {\bf Data from ScoF model}:      
        Table~\ref{ScoF} shows results on data generated from the ScoF(\textit{FR}) model. Both SI-ScoF(\textit{FR})and ScoF(\textit{FR}) show perfect $R^2$’s. The proposed model, SI-ScoSh, also shows near-perfect prediction. ScoSh fails to capture the predictor-response relationship when phases are not nuisances.

        From these experiments, we conclude that treating (predictor) phases as informative, when the data is generated using arbitrary phases, reduces the performance substantially. Conversely, ignoring the phases when they contain relevant information also impairs performance. Interestingly, the index function $h$ can compensate to some extent for phase mistreatment, making indexed models perform better than non-indexed ones. However, this compensation is limited to simpler $h_{true}$ and $\beta_{true}$; as they get more complex in shape, the index function struggles to compensate for phase mistreatment.

        \subsection{Evaluating Parameter Estimation}
        This section systematically evaluates estimation performances for different model parameters using simulated data.

\begin{figure}[!h]
            \centering
            \hspace{-1cm}
            \begin{tabular}{|cc|c|}
            \hline
                 \includegraphics[width=4cm]{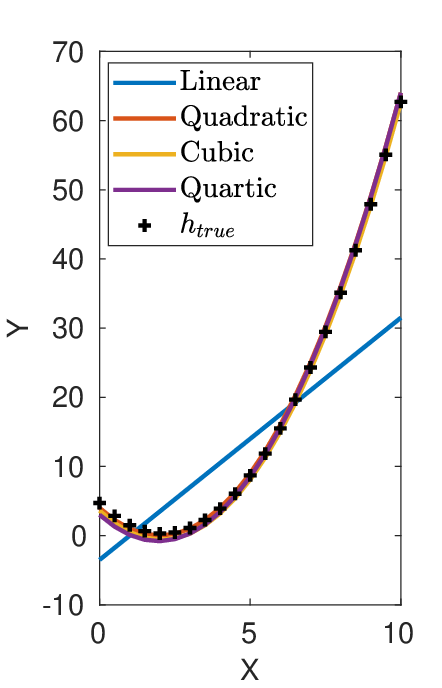}& 
                 \includegraphics[width=4.5cm]{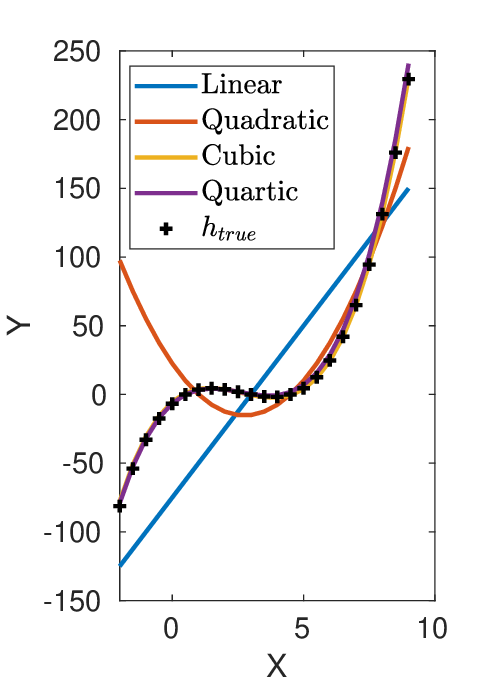}&
                 \includegraphics[width=5cm]{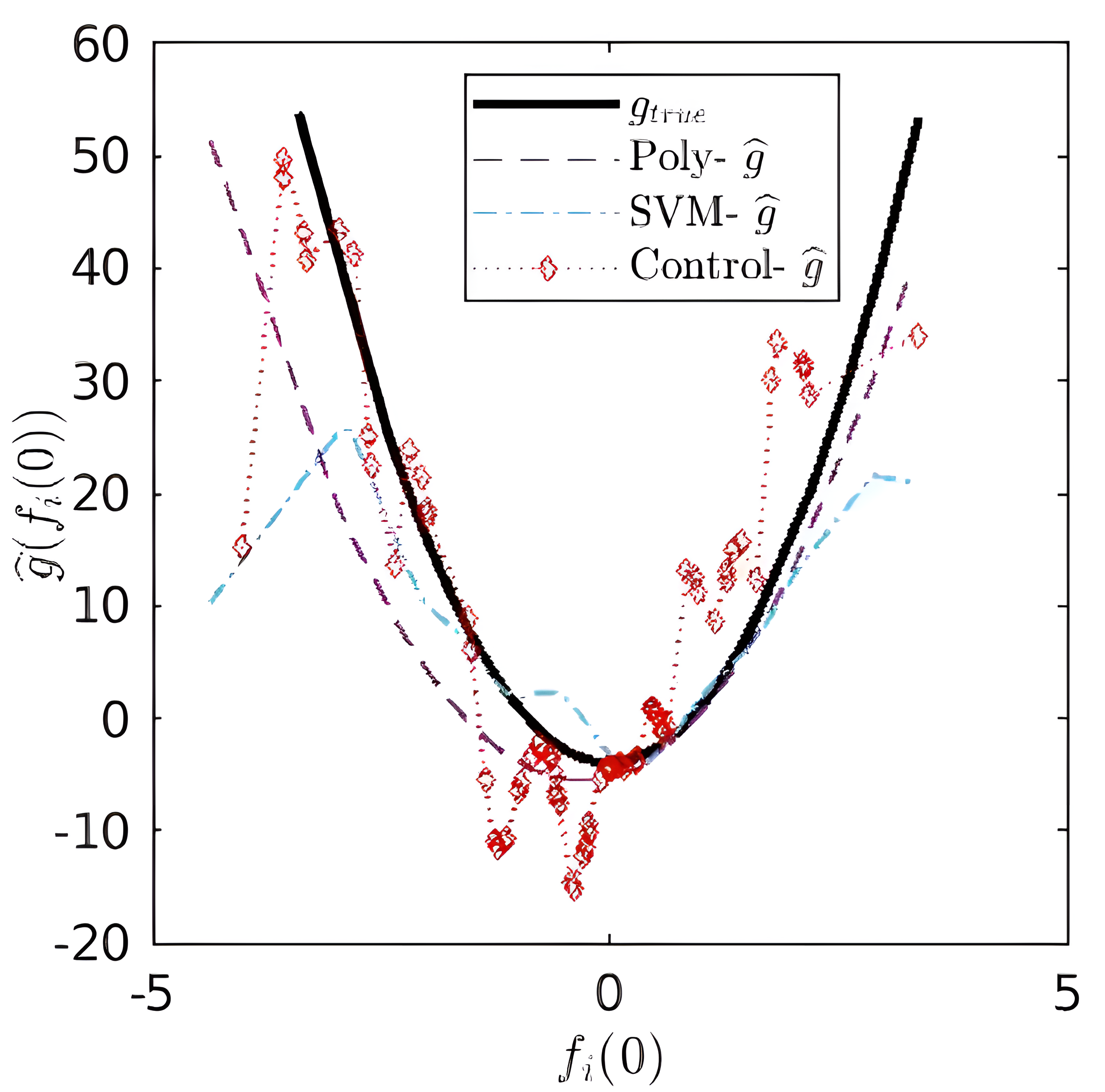}\\
                 \hline
            \end{tabular}

            \raggedright
            \caption{The estimate $\widehat{h}$ for $h_{true}=x^2-4x+4$ (left) and $h_{true}=(x-\frac{1}{2})(x-3)(x-4.5)$ (middle) under the SI-ScoSh model. The right panel shows estimates of $\widehat{g}$ with a quadratic $g_{true}$ for SI-ScoSh and SI-ScoF(\textit{FR})models.}
            \label{h's}
        \end{figure}

        \noindent 1. {\bf Estimation of Index Function $h$}: 
        In this experiment, we study how the varying degree of the index function $h$ affects the estimation performance of the SI-ScoSh model. We generate data from a quadratic or cubic $h_{true}$ and allow different degrees ($1-4$) during estimating of $h$. The pictorial results are shown in Fig.~\ref{h's} while error summaries are presented in Table~\ref{h1}. The left two panels of Fig.~\ref{h's} show estimated $h$ for different $h_{true}$. One can see that higher-order polynomials improve estimation. 

\begin{figure}[!h]
        \centering
            \begin{tabular}{|cc|c|}
            \hline
                 \includegraphics[width=4.5cm]{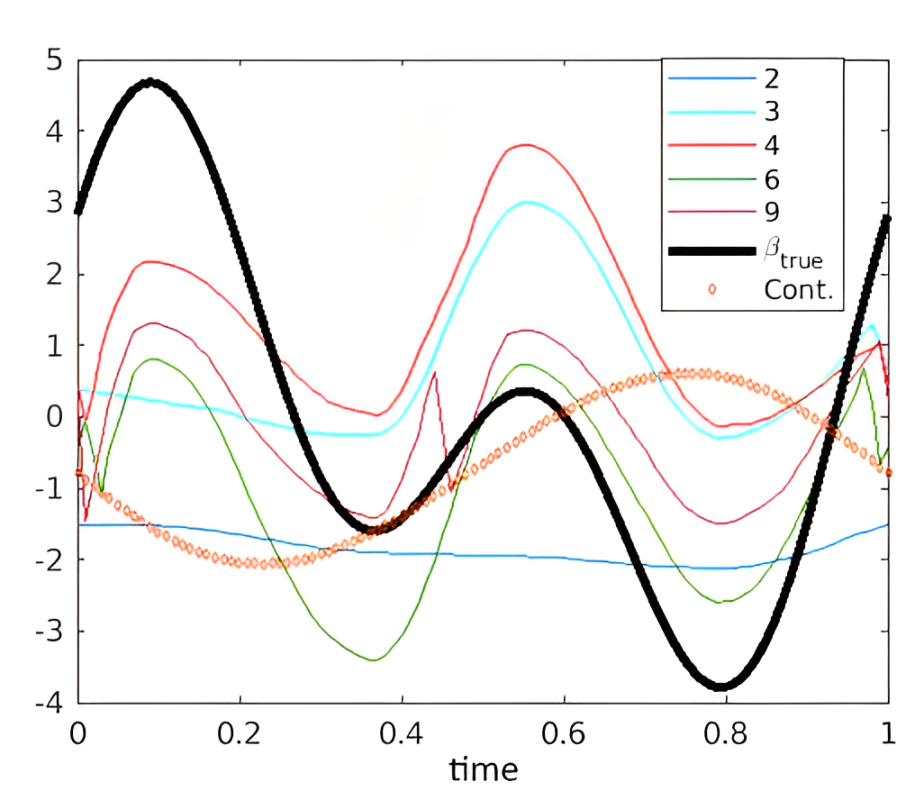}& 
                 \includegraphics[width=4.5cm]{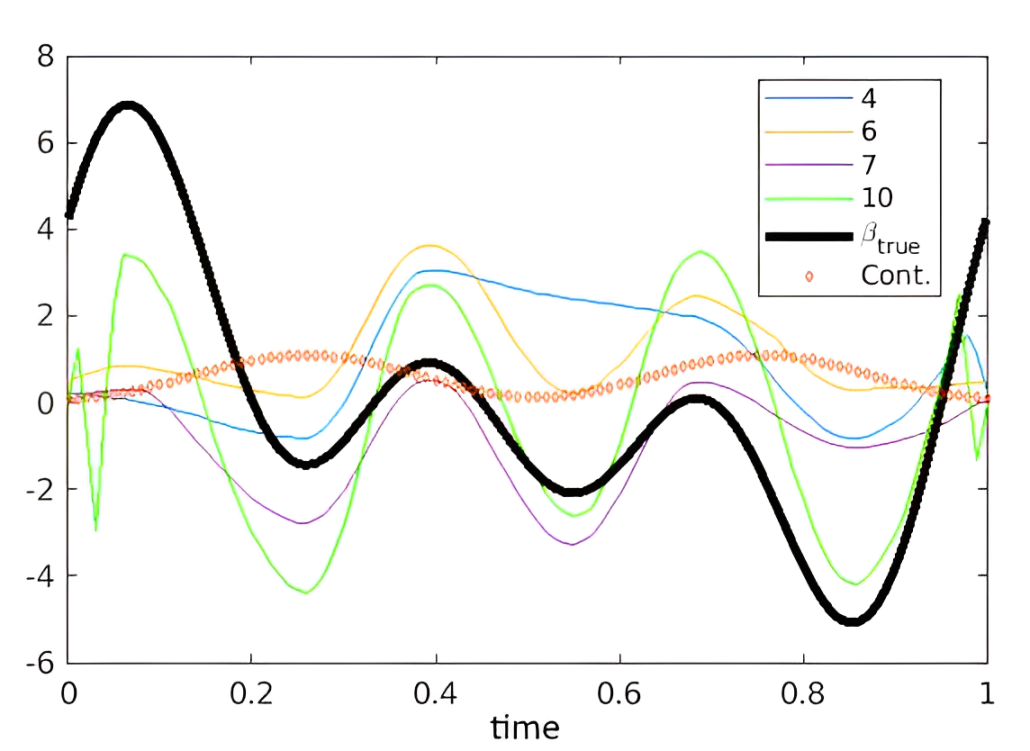}&
                 \includegraphics[width=5.cm]{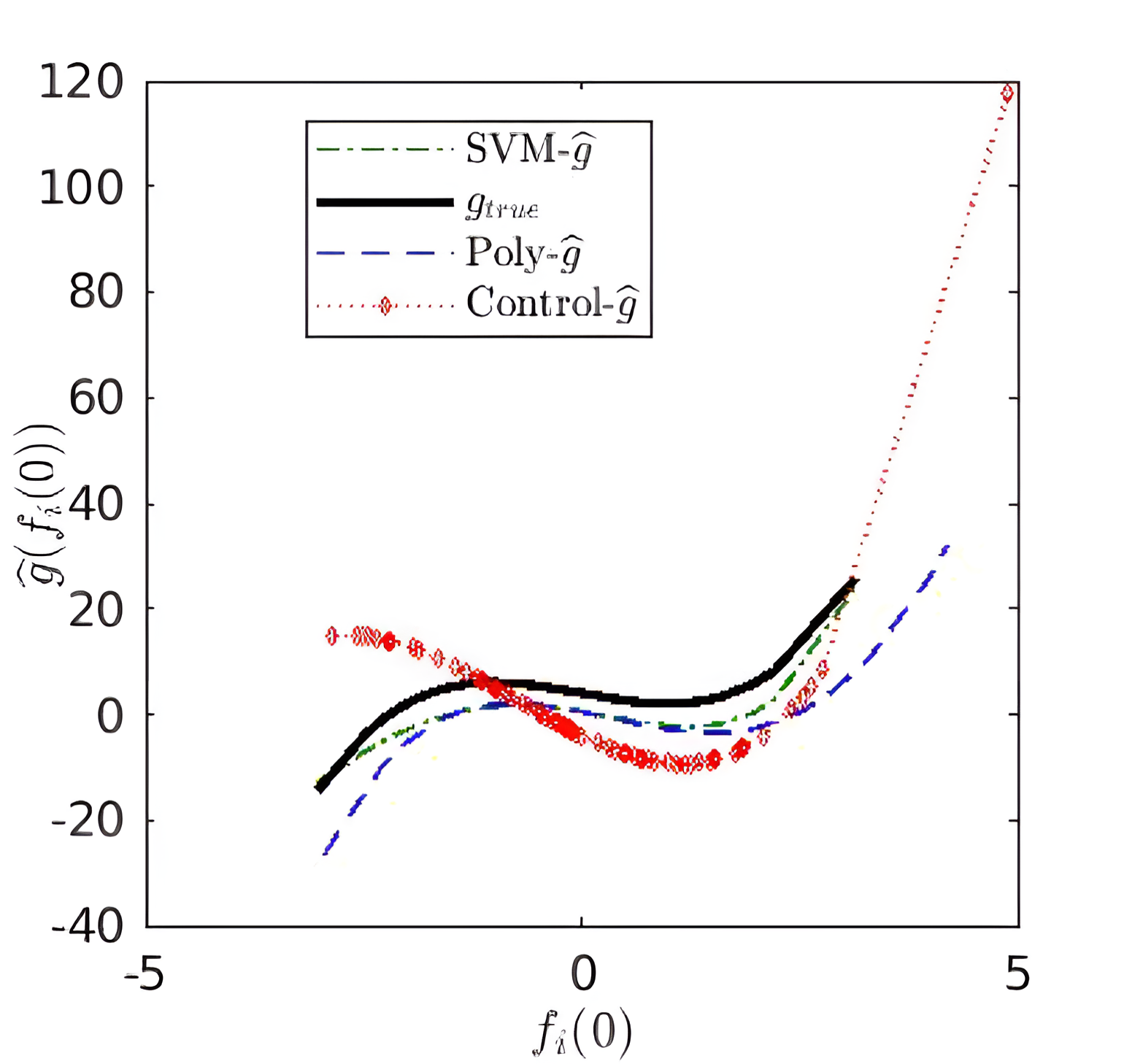} \\
                 \hline
            \end{tabular}

            \raggedright
            \caption{$\widehat{\beta}$'s when $\beta_{true}$ has $J=4$ (left) and  $J=6$ (middle). The numbers beside the colored lines in the legend show $J$ used in estimating $\beta$. The orange diamonds show the estimated $\beta$ for the SI-ScoF(\textit{FR}) model.  The right panel shows estimates of $\widehat{g}$ with a cubic $g_{true}$ for SI-ScoSh and SI-ScoF(\textit{FR}) models.}
            \label{beta's}
        \end{figure}

        \begin{table}[!h]
            \centering
            \caption{Prediction performance comparison for different complexities of $\widehat{h}$ with a quadratic (top) and cubic (bottom) $h_{true}$ having $\beta_{true}=\sum_{i=1}^{4}\sqrt{2}b_i(t)$ and $g_{true}(x)=x^2-1$.}
            \begin{tabular}{|c|c||c|c|c|c|c|}
                \hline
                $h_{true}$ & Pred. Performance & \multicolumn{4}{|c|}{SI-ScoSh : Maximum allowed degree of $\widehat{h}$} & SI-ScoF(\textit{FR})\\
                \hline\hline
                 \multirow{3}{*}{\rotatebox[origin=c]{90}{quadratic}} & Test $R^2$ & linear & quadratic & cubic & quartic &  \\
                 & Mean(SD) & $0.87(0.05)$ & $0.98(0.01)$ & $0.99(0.01)$ & $0.98(0.01)$ & $0.66(0.27)$\\
                 \cline{2-7}
                 & RMSE $(\widehat{h}-h_{true})$ & 4.15 & 0.13 & 0.19 & 0.20 & 9.1\\
                 \hline
                 \multirow{3}{*}{\rotatebox[origin=c]{90}{cubic}} & Test $R^2$ & linear & quadratic & cubic & quartic &  \\
                 & Mean(SD) & 0.67(0.11)& 0.74(0.22) & 0.95(0.04) & 0.96(0.03) & $0.51(0.2)$\\
                 \cline{2-7}
                 & RMSE $(\widehat{h}-h_{true})$ & 10.83& 6.0 & 2.3 & 3.5 & 24.8\\
                 \hline
            \end{tabular}
            \label{h1}
        \end{table}  
\noindent        
    2. {\bf Estimation Regression Coefficient $\beta$}: Here we study estimation of $\beta$  using different basis sets of $\ltwo$ space. We construct $\beta_{true}$ from $4$ or $6$ Fourier basis elements, and we estimate it under the SI-ScoSh model for different $J$ values. As seen in the left and middle panels of Fig.~\ref{beta's}, increasing $J$ beyond the true degrees doesn't improve estimation of $\widehat{\beta}$ any further. 
    This trend is mirrored in the predictive $R^2$'s, presented in Table~\ref{beta6}. For $J<J_{true}$, $\widehat{\beta}$'s have worse prediction performance compared to $J\ge J_{true}$ as they fail to capture the shape of the $\beta_{true}$. 
     Fig.~\ref{beta's} and Table~\ref{beta6} show that further increasing the number of basis elements for $\beta$ does not necessarily improve performance.

    Note that we use the shape metric $d_s$, rather than RMSE, for evaluating $\widehat{\beta}$. As discussed in Section 2.2, the shape of $\beta$ is more relevant in ScoSh model than $\beta$ itself.

    \begin{table}[!h]
    \centering\small
    \caption{Prediction performance comparison for different complexities of $\widehat{\beta}$ with $\beta_{true}=\sum\limits_{i=1,3}2b_i+\sum\limits_{i=2,4}\sqrt{2}b_i$ (top) and $\beta_{true}=\sum\limits_{i=1,3,5}2b_i+\sum\limits_{i=2,4,6}\sqrt{2}b_i$ (bottom). The used $h_{true}(x)=x^2-3x+2$ and $g(x)=x^2-1$ }
        \begin{tabular}{|c|c||c|c|c|c|c|c|}
            \hline
            $\beta_{true}$ & Pred. Performance & \multicolumn{5}{|c|}{SI-ScoSh : Number of basis functions for $\widehat{\beta}$} & SI-ScoF(\textit{FR})\\\hline\hline
            \multirow{3}{*}{\rotatebox[origin=c]{90}{J=4}} & Test $R^2$ & J=2 & J=3 & J=4  & J=6 & J=9 &  \\
             & Mean(SD) & $0.75(.07)$ & $0.93(.03)$ & $0.99(.01)$ & $0.98(.01)$ & $0.99(.01)$ & $0.62(.14)$\\
             \cline{2-8}
             & RMSE $(\widehat{\beta}-\beta_{true})$ & 3.6 & 2.8 & 2.4 & 3.8 & 3.2 & 5.6 \\
            \hline
            \multirow{3}{*}{\rotatebox[origin=c]{90}{J=6}} & Test $R^2$ & \multicolumn{5}{|c|}{\multirow{3}{*}{\begin{tabular}{c|c|c|c}
                 J=4 & J=6 & J=7 & J=10   \\
                 $0.85(0.08)$ & $0.96(0.02)$ & $0.99(0.01)$ & $0.99(0.01)$  \\
                 4.9 & 4.0 & 3.8 & 4.5
             \end{tabular}}} & \\
             & Mean(SD) & \multicolumn{5}{|c|}{} & $0.4(0.45)$ \\
             \cline{2-8}
             & RMSE $(\widehat{\beta} -\beta_{true})$ & \multicolumn{5}{|c|}{} & 7.2\\
             \hline
        \end{tabular}
        
        \label{beta6}
    \end{table}
        
\noindent 3. {\bf Estimation Error for $g$}:
        Here, data is generated with a fixed $h$ (a quadratic) and $\beta$ composed of four Fourier basis elements, but we set $g_{true}$ to be either quadratic or cubic. Then, we estimate $g$ under the SI-ScoSh model and see how well we recover the structure of $g_{true}$ under different models. The results are shown in the rightmost panels of Figs.~\ref{h's} and \ref{beta's}.  Both the relative RMSE between the true and estimated $g$ and the prediction performances (see Table~\ref{gg}) establish the superiority of the SI-ScoSh model over the SI-ScoF(\textit{FR}) model.

        \begin{table}[!h]
                \centering
                \small
                    \caption{Prediction performance comparison among different models for different $g_{true}$'s}
                    \begin{tabular}{|cc|cc|c||c|c|c|c|}
                        \hline
                        \multirow{5}{*}{\rotatebox[origin=c]{90}{$h_{true}(x)=$}} & \multirow{5}{*}{\rotatebox[origin=c]{90}{$x^2-3x+2$}} & \multirow{5}{*}{\rotatebox[origin=c]{90}{$\beta_{true}=\sum\limits_{i=1,3}2b_i(t)+$}} & \multirow{5}{*}{\rotatebox[origin=c]{90}{$\sum\limits_{i=2,4}\sqrt{2}b_i(t)$}} &$g_{true}(x)$ & \multicolumn{2}{|c|}{$x^3-3x+4$} & \multicolumn{2}{|c|}{$5x^2-4$}\\
                        \cline{5-9}
                        &&&& Prediction & $R^2$ & $\frac{||\widehat{g}-g_{true}||}{||g_{true}||}$ & $R^2$ & $\frac{||\widehat{g}-g_{true}||}{||g_{true}||}$\\
                        \cline{5-9}\cline{5-9}
                        &&&& SI-ScoSh (Poly) & 0.99 & 0.33 & 0.98 & 0.36 \\ 
                        &&&& SI-ScoSh (SVM) & 0.98 & 0.28 & 0.98 & 0.29\\
                        &&&& SI-ScoF(\textit{FR}) & 0.54 & 0.59 & 0.12 & 0.97 \\
                        \hline
                    \end{tabular}  
                    \label{gg}
            \end{table}

\subsection{Evaluating Model Invariance to Random Phases} 
    The main goal of this paper is to design a regression model that is invariant to phase variability in predictor functions. While the proposed ScoSh and SI-ScoSh models satisfy this requirement theoretically, we also evaluate this property empirically. Specifically, we design response variables $y_i$s that are by definition invariant to phase shifts in $f_i$. In other words, the responses are dependent exclusively on the shape of the corresponding predictor.  We choose two cases: (1)  $y_i= (\max(f_i(t))-\min(f_i(t)) + \epsilon_i$, (2) $y_i=\int\limits_0^1|\dot{f_i}(t)|~dt + \epsilon_i$.
Here $\epsilon_i\sim\mathcal{N}(0,0.5)$,
and the predictors $\{f_i\}$ are generated as in Section~3. Then, we apply the proposed model to the noisy and time-warped data and study the results.

\begin{figure}[!h]
    \begin{tabular}{cccc}
        \includegraphics[width=5cm]{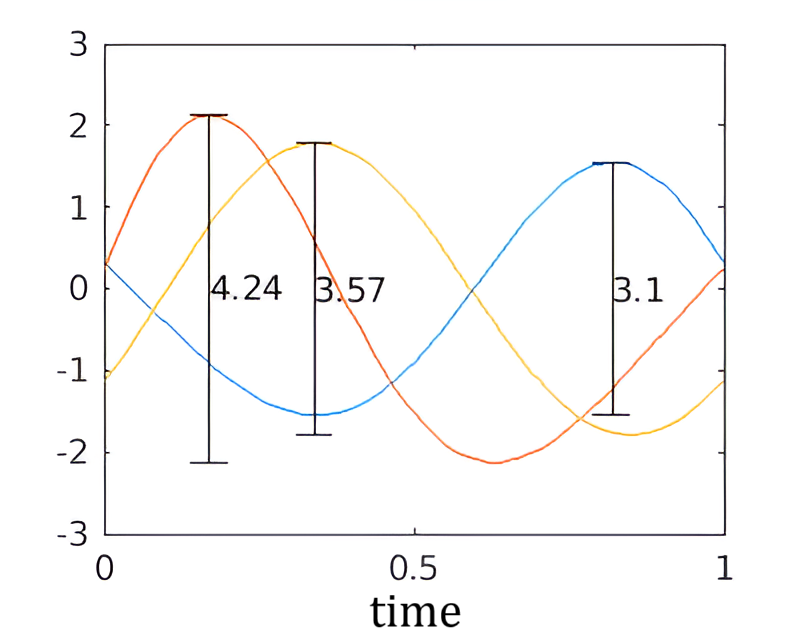}
        \hspace{-0.2cm}\includegraphics[width=3.9cm,height=4.05cm]{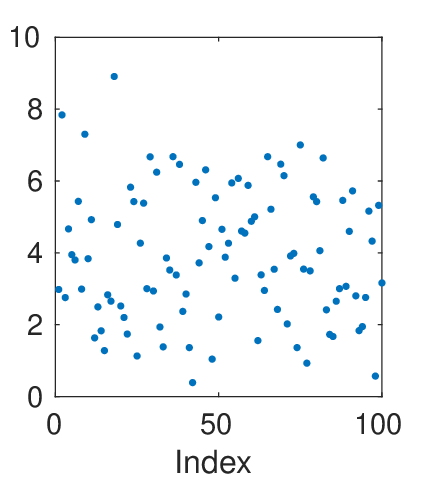} &
        \hspace{-0.5cm}\includegraphics[width=3.9cm]{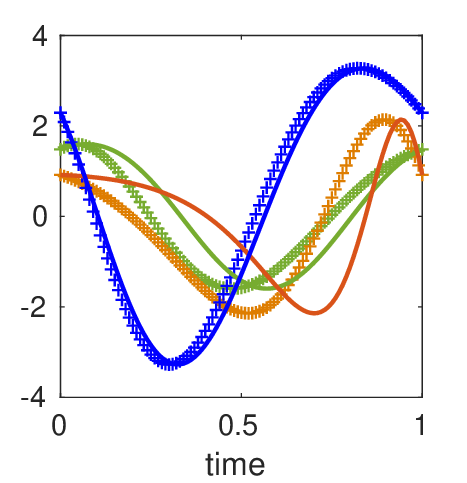} &
        \hspace{-0.5cm}\includegraphics[width=3.7cm]{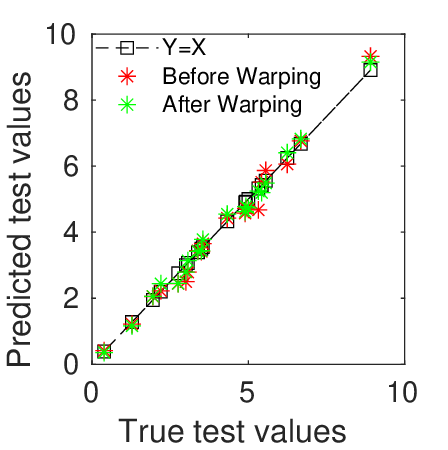} \\
        \hspace{-3.9cm}\includegraphics[width=4.3cm]{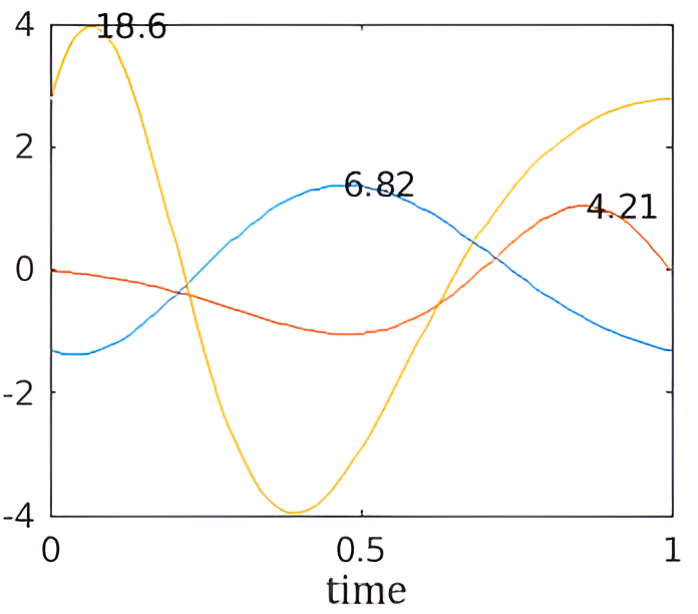} & 
        \hspace{-8.2cm}\includegraphics[width=3.9cm]{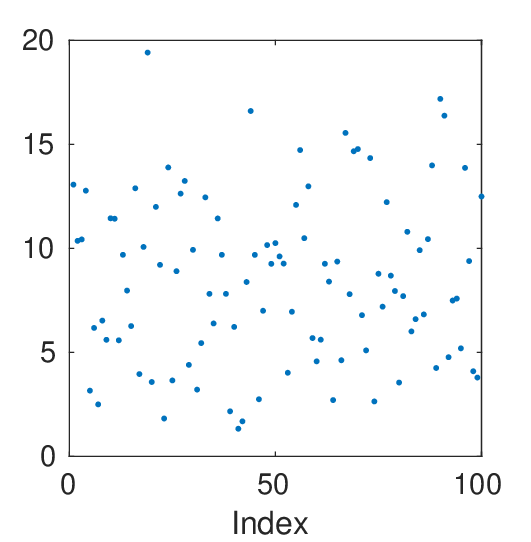} &
        \hspace{-7.9cm}\includegraphics[width=4.04cm]{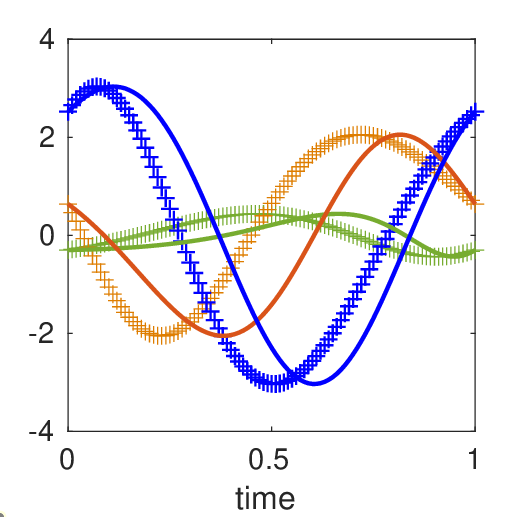} & 
        \hspace{-4.1cm}\includegraphics[width=3.7cm]{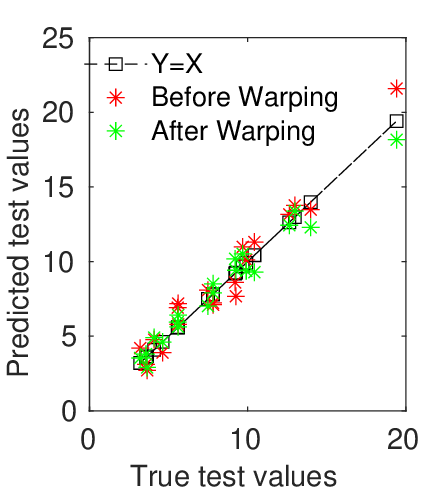}
    \end{tabular}
    \caption{Examples of $f_i$s (left), noisy measurements of $y_i$s (second), the original test set in solid lines and their perturbed version in hashed lines (third panel), and SI-ScoSh predicted $\widehat{y}_i$ plotted versus true $y_i$s (last panel). Top: Responses $y_i$ are the max amplitude of predictors $f_i$.  Bottom: Responses $y_i$ are lengths of predictors $f_i$.}
    \label{Sres1}
\end{figure}

Fig.~\ref{Sres1} presents results from these experiments. 
The two rows show results for two data cases. We train the models with a training set and evaluate them on a separate test set. Finally, we compare the prediction performances of SI-ScoSh and SI-ScoF(\textit{FR}) on test sets. SI-ScoSh achieves $R^2 = 0.98$, while SI-ScoF(\textit{FR}) has $R^2 < 0.1$. SI-ScoF(\textit{FR})'s lack of optimization over $\gamma_i$s results in inferior performance. The high performance of the ScoSch model underscores its invariance to random phases of predictor functions.

 \section{Experimental Results: Real Data}

In this section, we investigate the use of proposed ScoSh models on several real datasets. In each case, the functions are given without any prior registration, and we investigate the effectiveness of regressing scalar responses on the shapes of predictors. The detailed prediction performances of the different models are provided in a table format in the \textcolor{blue}{Supplementary Material}.
 
 \begin{enumerate}[leftmargin=*]
\item {\bf Spanish Weather Data}: This data contains daily summaries of geographical data of 73 Spanish weather stations selected from 1980-2009. Although this dataset contains other variables measured at each weather station, we focus only on the temperatures.
We form a predictor function $f_i$ for each station with 365 average temperature values as follows. Each value is the average temperature recorded on a day ({\it e.g.}, February $3^{rd}$) for all years from 1980 to 1993. The corresponding scalar response $y_i$ is the mean of temperatures for all days between 1994 and 2009 at that station. This data is shown in the top row of Fig.~\ref{SWD}. The goal is to use past temperature patterns for each station to predict future average temperatures. 

         \begin{figure}[!h]
            \centering
            \begin{tabular}{ccc}
                \includegraphics[width=5cm]{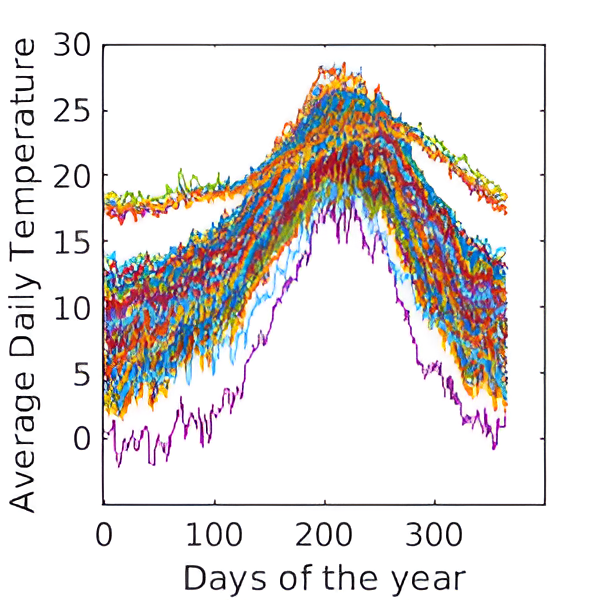} \vspace{-0.25cm}&
                \includegraphics[width=4.7cm]{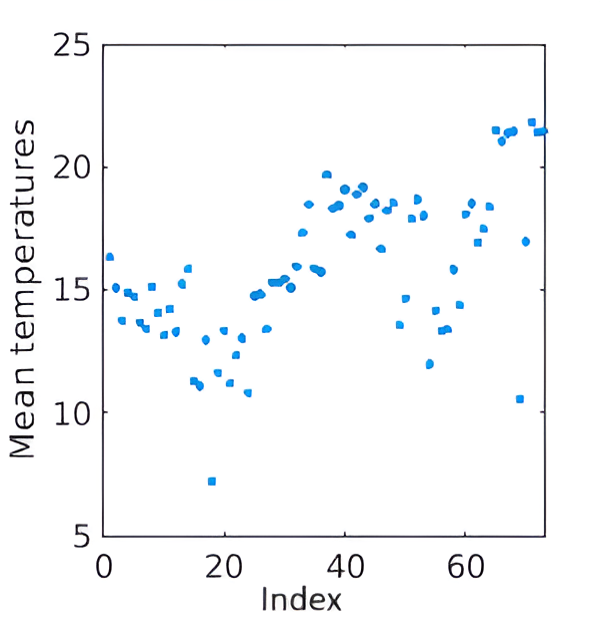} &
                \includegraphics[width=5.7cm]{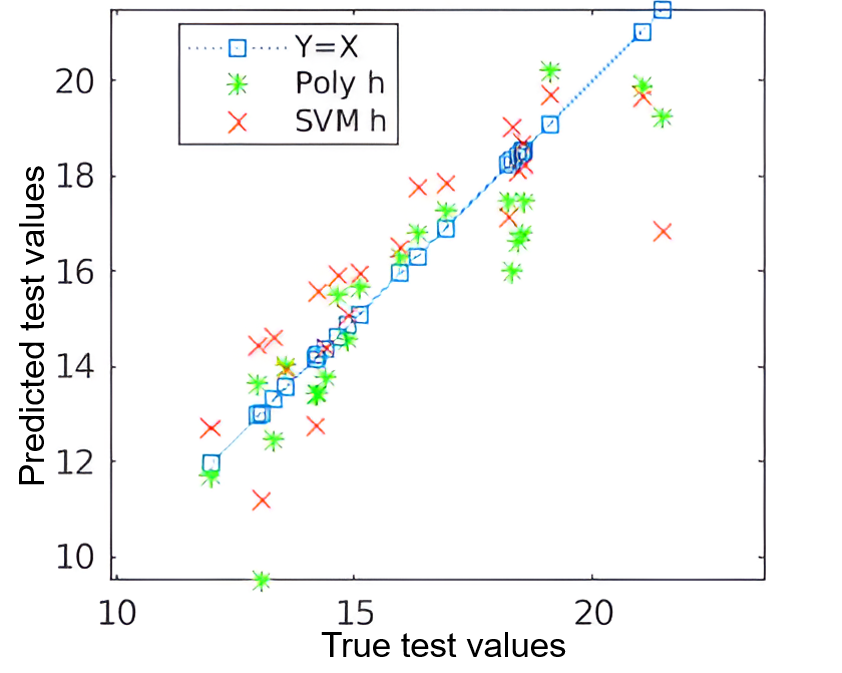}\\
                \includegraphics[width=5cm]{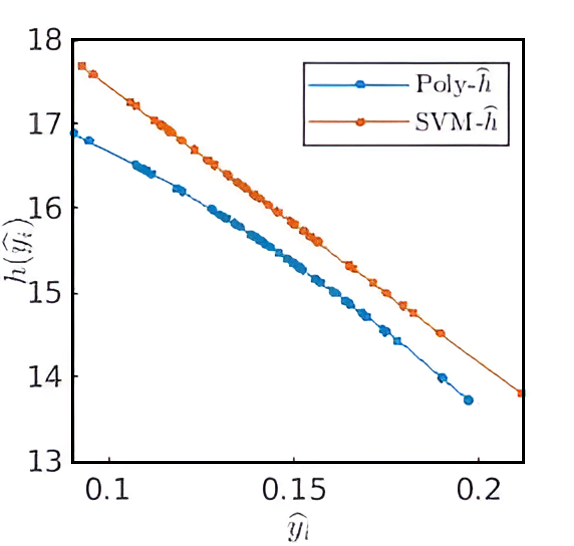}\vspace{-0.5cm}&
                \includegraphics[width=5cm]{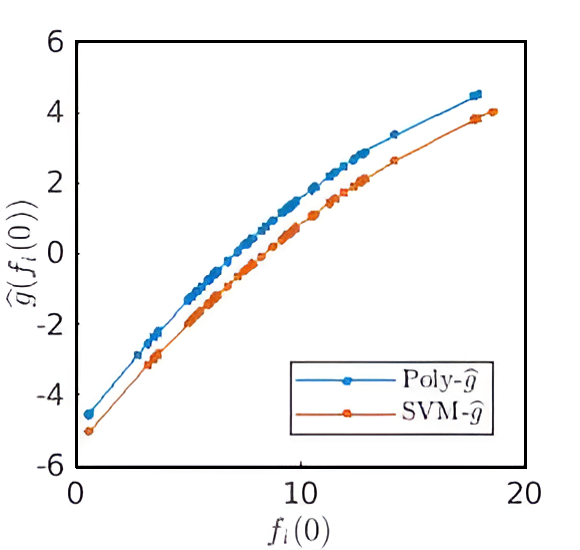}& 
                \includegraphics[width=5cm]{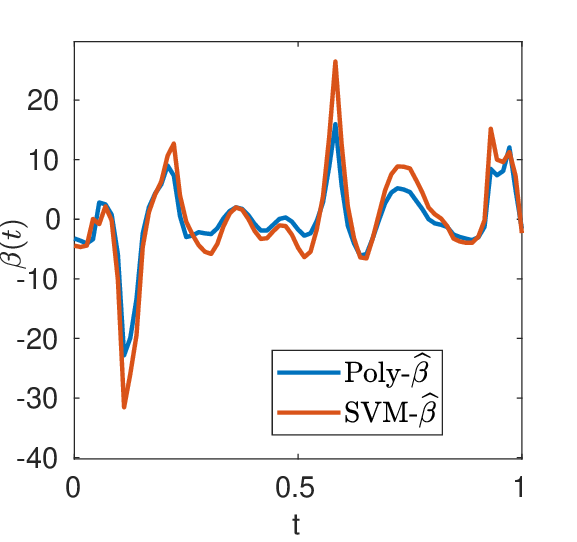}   
            \end{tabular}
            \caption{Spanish weather results -- Top:  Predictor functions $\{f_i\}$ (left), the responses $\{y_i\}$ (middle), and model predictions $\{\widehat{y}_i\}$ against true test values $\{y_i\}$ (right).
            Bottom: the estimated parameters under SI-ScoSh model -  $\widehat{h}$ (left), $\widehat{g}$ (middle), and $\widehat{\beta}$ (right).}
            \label{SWD}
        \end{figure}

We apply the proposed and the competing models to this dataset to evaluate their prediction performances. We use two versions of SI-ScoSh. For estimating index function $h$ with a parametric curve, we obtain $R^2 = 0.92$ on the test set, and for the non-parametric method (SVM) with different kernels (Polynomial/Rbf), we get $R^2 = 0.89$. SI-ScoSh performs best among all models, while, in contrast, SI-ScoF($FR$) gives a prediction performance of merely $R^2=0.58$ and SI-ScoF($\ltwo$) an $R^2=0.45$. Simpler indexed models fail to capture these relationships -- $R^2$'s are less than $0.1$ for ScoF ($\ltwo$ \& $FR$) and ScoSh. The parameter estimates of the SI-ScoSh model are shown in Fig.~\ref{SWD}.

 \item {\bf Covid Hospitalization Data}:
        This data\footnote{https://ourworldindata.org/covid-deaths} contains the number of daily new COVID hospitalizations at hospitals in 31 European countries, which serve as our predictors. 
        The observation period is from January 1, 2020, to October 13, 2022, so each predictor $f_i$ contains 1016 elements. The responses $y_i$ are the total number of deaths in the respective countries that occurred during the observation period. Our goal is to utilize these hospitalization curves to predict the number of fatalities in a country. The data and the results are presented in Fig.~\ref{CHD1}.

        \begin{figure}[!h]
                \centering\small
            \begin{tabular}{ccc}
                \includegraphics[width=5.2cm]{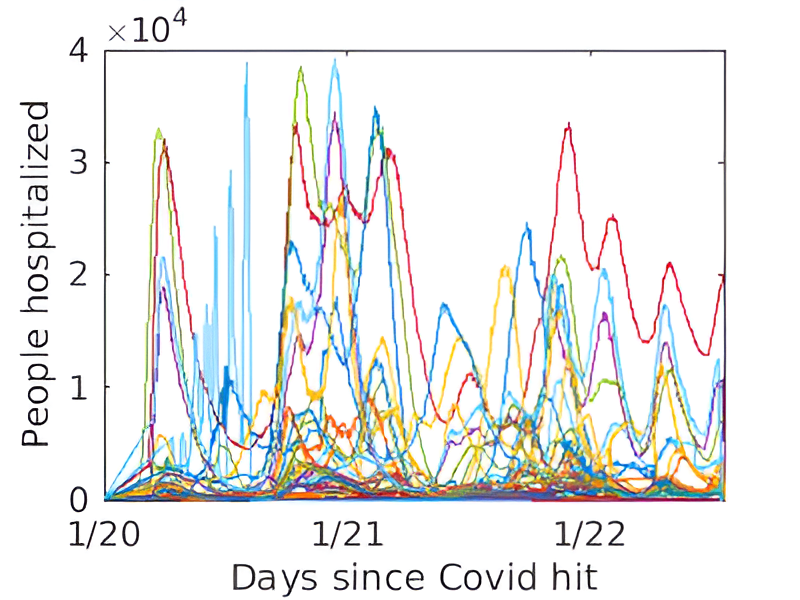} &
                \includegraphics[width=5cm]{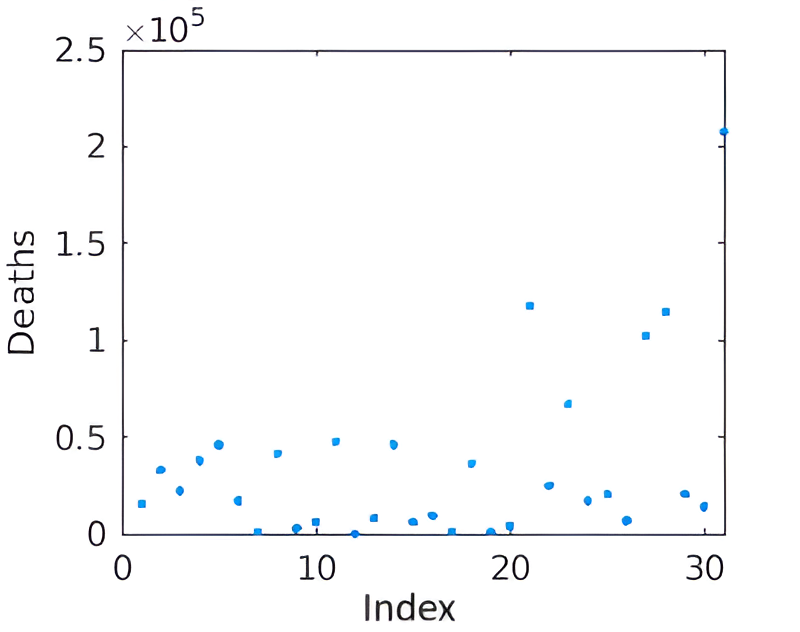} &
                \includegraphics[width=4cm]{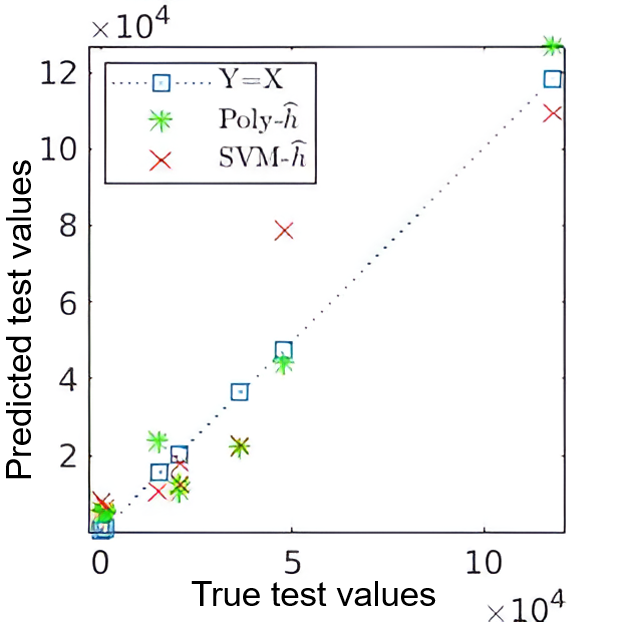}\\
                \includegraphics[width=5cm]{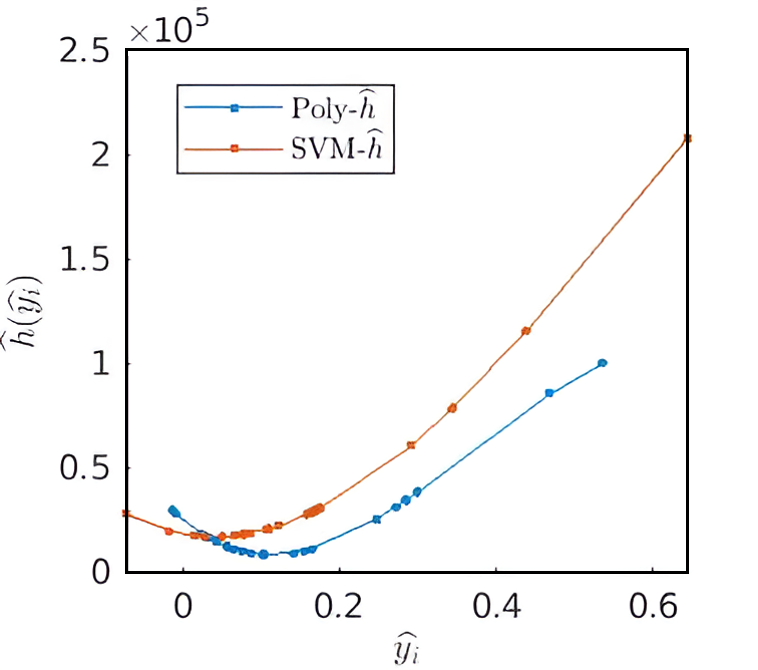}\vspace{-0.5cm}&
                \includegraphics[width=5cm]{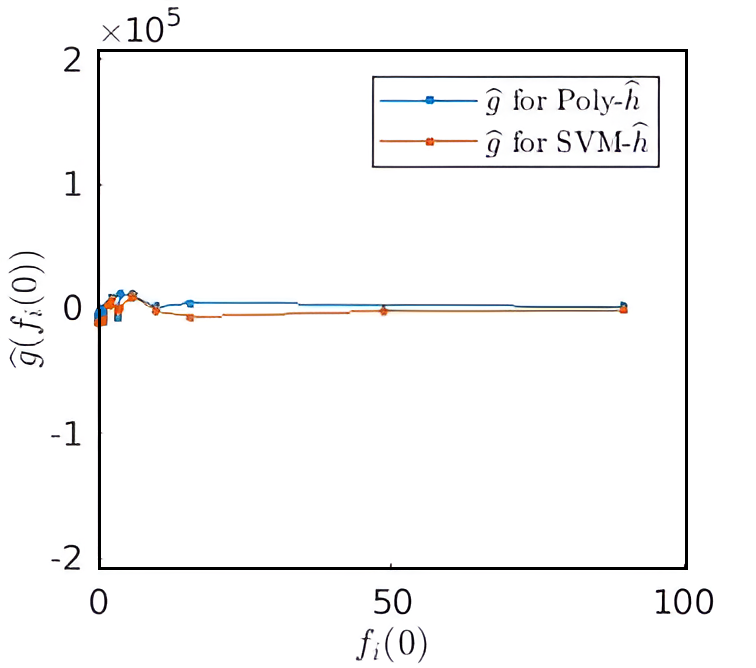}&
                \includegraphics[width=4.8cm]{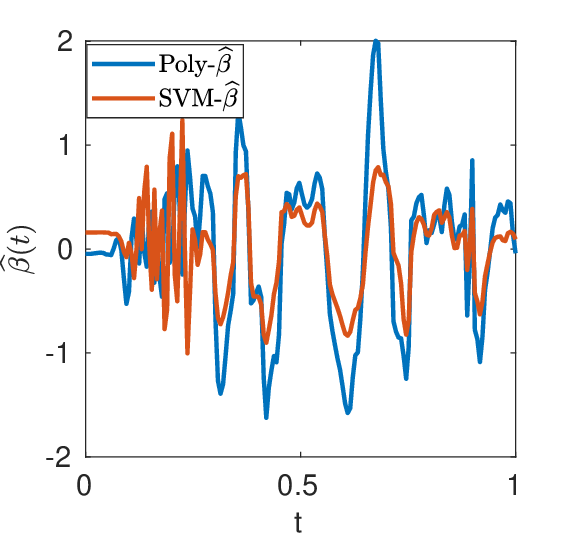}
            \end{tabular}
            \caption{Covid hospitalization results -- Top: the daily hospitalization curves (left), corresponding fatality counts (middle), and predicted responses versus true responses (right). Bottom: the estimated parameters under SI-ScoSh -- $\widehat{h}$ (left), $\widehat{g}$ (middle), and $\widehat{\beta}$ (right).}
            \label{CHD1}
        \end{figure}

 Under the SI-ScoSh model -- a quadratic $g$, a cubic $h$, and a $\beta$ using the first six Fourier basis elements -- provide the best performance (test set prediction $R^2>0.92$).  The estimates in the bottom of Fig.~\ref{CHD1} show that $\widehat{g}$ is relatively constant when compared to $\widehat{h}$, indicating most correlation is captured by $\beta$ and $h$. This is because all countries start from a point of zero hospitalizations, {\it i.e.}, $\{f_i(0)\}$ are all zero. Other models like SI-ScoF ($\ltwo$ \& $FR$), ScoSh and ScoF ($\ltwo$ \& $FR$) fail to capture significant relationships with prediction $R^2$'s less than $0.2$. For details, please refer to the \textcolor{blue}{Supplementary Material}.

 \item {\bf Covid Infection Data}:
        This dataset\footnote{https://ourworldindata.org/covid-hospitalizations}
        contains the number of new COVID-19 infections per day in each of 41 countries. These daily infection rate functions serve as the predictors. (see top left of Fig \ref{CID1}). The total number of people hospitalized during the entire period is the response for each country. The raw dataset has been smoothed but not centered or phase-shifted.
        

        \begin{figure}[!h]
            \begin{tabular}{ccc}
                 \hspace{-0.5cm}\includegraphics[width=6.2cm]{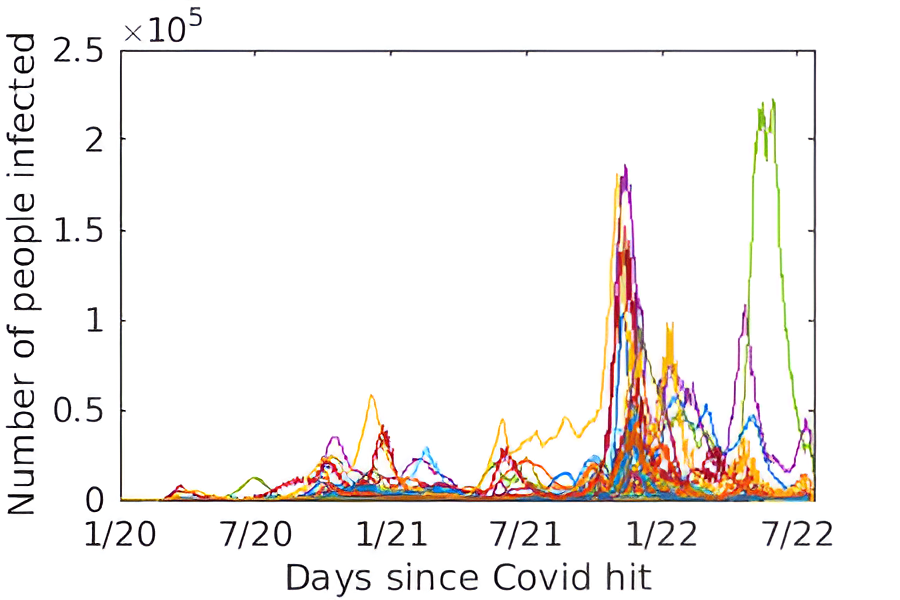}\vspace{-0.3cm}& 
                 \includegraphics[width=4.5cm]{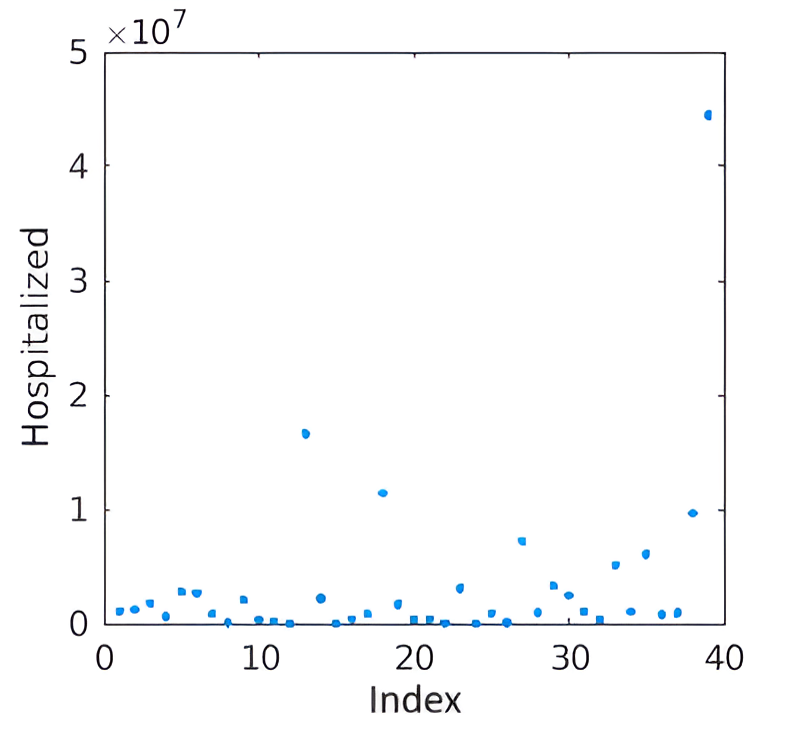}&
                 \hspace{-0.5cm}\includegraphics[width=5.2cm]{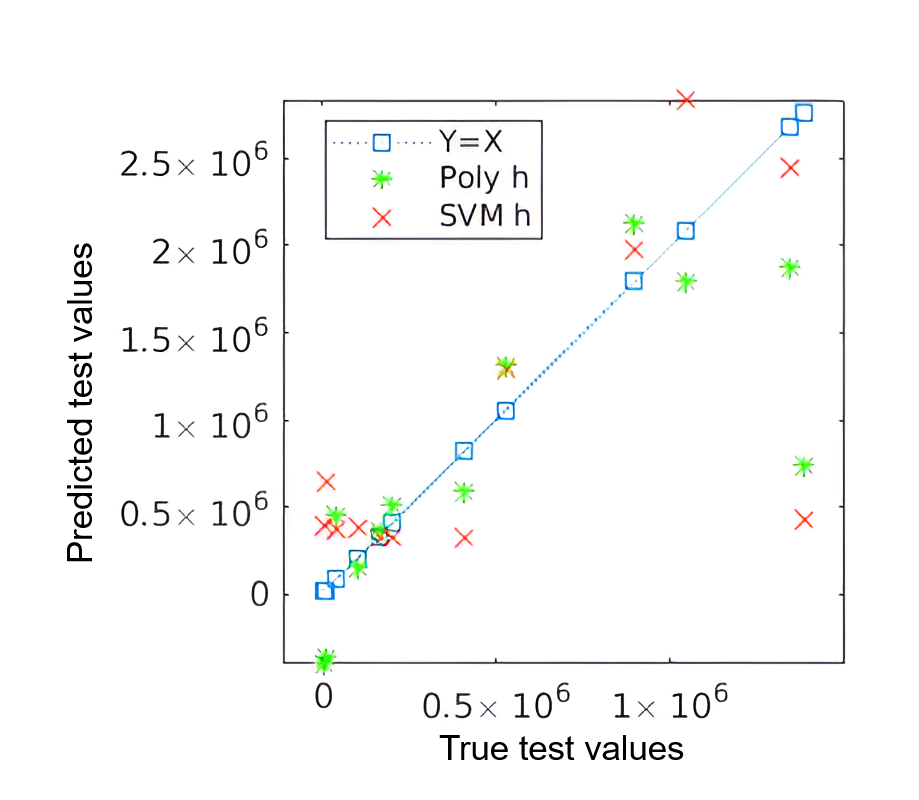}\\
                 \includegraphics[width=5cm]{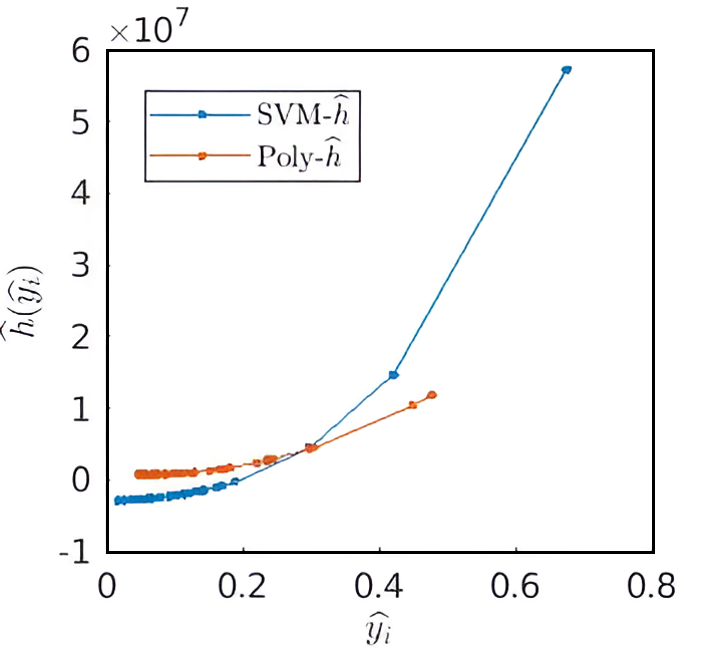}\vspace{-0.5cm}&
                 \includegraphics[width=5cm]{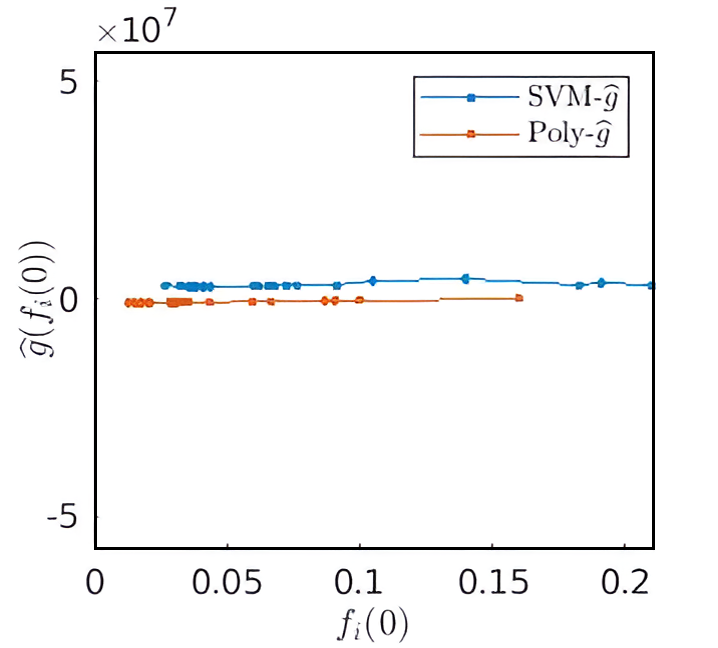}&
                 \includegraphics[width=5cm]{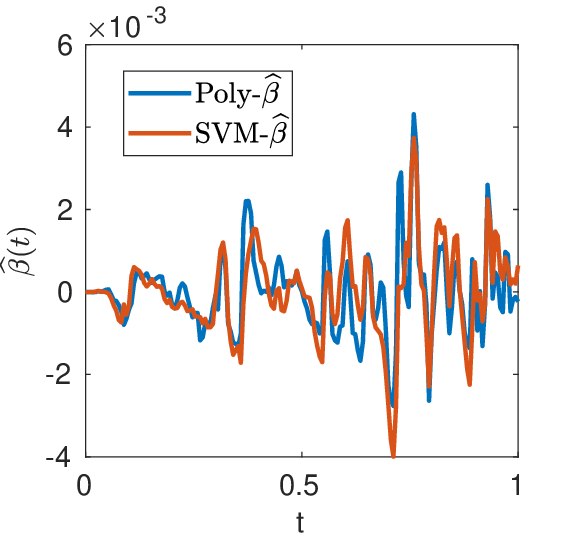}
            \end{tabular}
            
            
            \caption{Covid infection results -- Top: the daily infection curves (left), corresponding hospitalization counts (middle), and predicted responses versus true responses (right). Bottom: the estimated parameters under SI-ScoSh -- $\widehat{h}$ (left), $\widehat{g}$ (middle), and $\widehat{\beta}$ (right).
            }
            \label{CID1}
        \end{figure}

        We apply the SI-ScoSh and SI-ScoF(\textit{FR} \& $\ltwo$) models and their simpler versions for a prediction performance comparison. The SI-ScoSh model predicts the test responses with $R^2=0.89$,but the SI-ScoF(\textit{FR}) model captures a far less statistically significant predictor-response relationship ($R^2=0.4$). Models without the index functions provide even worse performance. The $\ltwo$ version of the ScoF model performs worse than the \textit{FR} version. Like the previous example, $g$, the index function does not play an important role here. Please refer to the \textcolor{blue}{Supplementary Material} for detailed results.

\end{enumerate}

\section{Extension to a Multiple Index Model}

Following \cite{F2013}, we can extend the SI-ScoSh model from a single index to a multiple index model according to: 
\begin{equation}\label{eq:Ferraty}
     y_i=  \sum_{j=1}^K \left\{ g_j(f_i(0))+ h_j\left(\sup_{\gamma_{i,j}\in\Gamma}{\left\langle\beta_j,q_i\star\gamma_{i,j}\right\rangle}\right) +\epsilon_{ij} \right\}\ ,\  i=1,\cdots,n
\end{equation}
The estimation proceeds by treating the problem as a single-index model and estimating $\{\beta_1,h_1, g\}$. Then, we calculate the residuals and use them as responses for the next single index model, leading to the estimation of $\{\beta_2,h_2, g_2\}$. We continue until the improvement in prediction performance becomes small. 

\begin{figure}[!h]
    \centering
    \begin{tabular}{cc}
         \includegraphics[width=5.5cm]{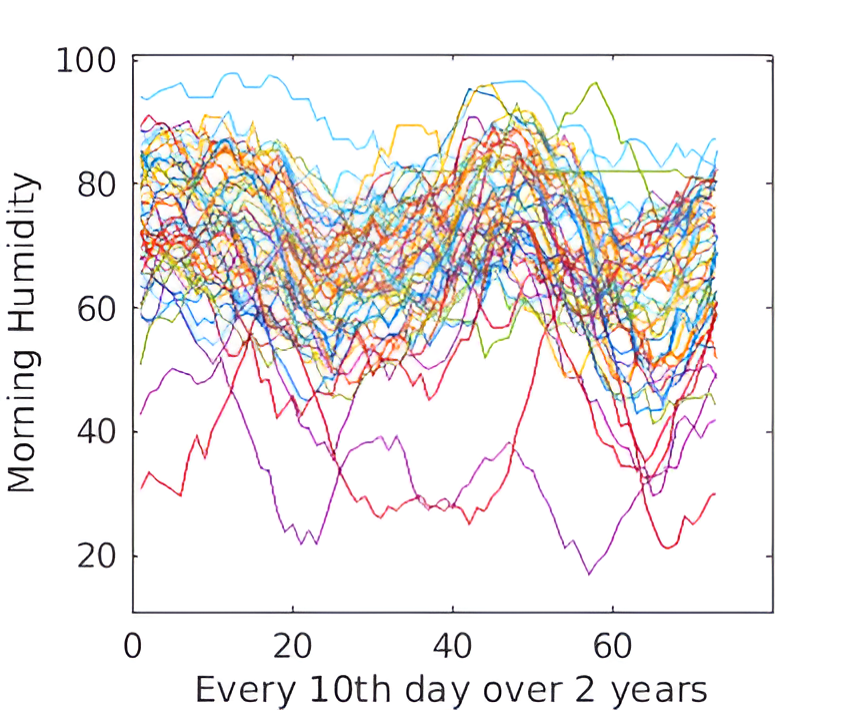}& 
         \includegraphics[width=5.5cm]{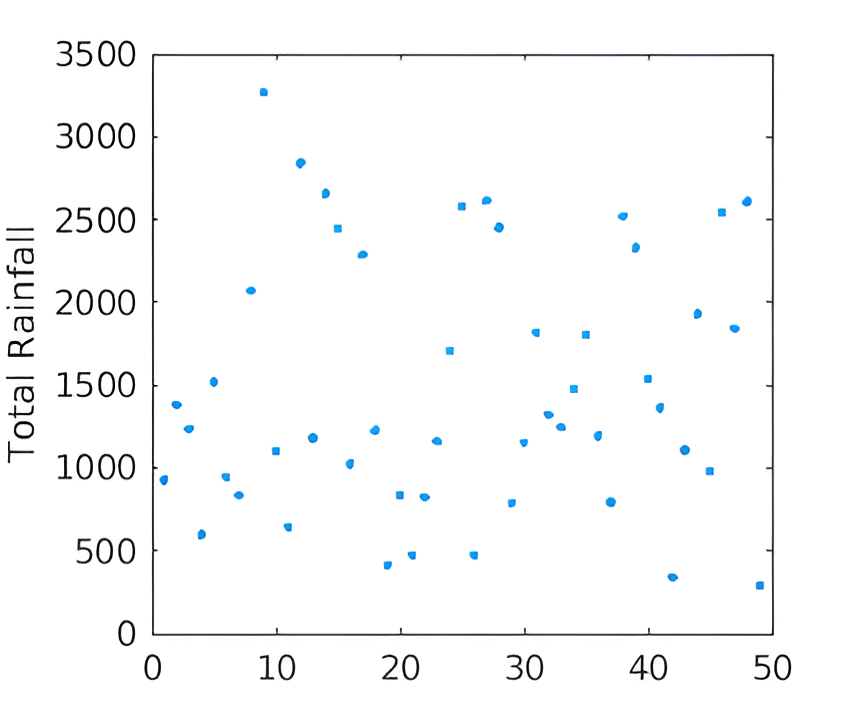}
    \end{tabular}
    \includegraphics[width=16cm]{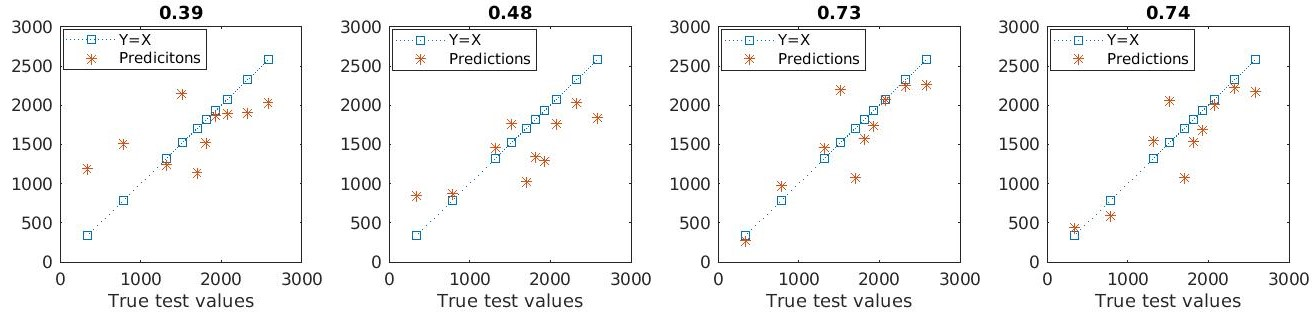}
    \caption{Top: the morning humidity curves for different counties (left) and the total rainfall in the respective counties (right). Bottom: Results from successive index models with $R^2$ values on top. These values improve from $0.39$ for $K=1$ to $0.74$ for $K=4$ .}
    \label{Ferrpred}
\end{figure}

\noindent \textbf{Rainfall vs Morning Humidity}: We illustrate this model using a weather dataset. 
The predictor functions are daily humidity at 9 am every ten days over the course of the period Jan-1-2014 to Dec-31-2015 for 49 counties in Australia. The response variable for each county is the total amount of rain over the same period.
 The raw dataset\footnote{https://rattle.togaware.com/weatherAUS.csv} has been smoothed (with a moving average) to reduce noise. The results from the application of the multi-index ScoSch model are presented in Fig.~\ref{Ferrpred}.

 The results show that the first layer $\{h_1,\beta_1,g_1\}$ captures approximately a third of the correlation between shapes of the predictors and the response, but as we add more layers, the prediction performance $R^2$ increases to around $0.74$. Further addition of layers does not improve performance. This result contrasts SI-ScoF(\textit{FR}), where $R^2$ improves less than $0.1$ for each extra layer. A detailed table is presented in the \textcolor{blue}{Supplementary Material}.

\section{Conclusion}
   
Functional data has two components: phase and shape, and they may contribute at different levels in a functional regression model. This paper develops a novel approach, termed a ScoSh model, that uses only the shapes of functions and ignores their original phases in scalar predictions. Furthermore, it optimizes the phases inside the regression models rather than as preprocessing, as is often done currently. This formalization leads to new definitions of \textit{regression phase} and {\it regression mean}. The model also imposes an index function to result in a SI-ScoSh model. The two novel components - removal of dependence on the predictor phase and using a nonlinear index function - show improved performance in various situations. Several simulated and real-data experiments demonstrate the model and its superiority. 

The proposed SI-ScoSh model is appropriate when the phase components of predictors carry little or no information. This is often the case in image analysis and neuroimaging, where phases correspond to different parameterizations of neuroanatomical objects. However, in general, the phase components may contain helpful information, and discarding them would degrade prediction performance. In that situation, a more flexible model would be to separate the phases (from shapes) and use them as separate predictors themselves. This idea has been left for future explorations.

\bibliographystyle{abbrvnat}

\bibliography{biblio}

\begin{thebibliography}{37}
\providecommand{\natexlab}[1]{#1}
\providecommand{\url}[1]{\texttt{#1}}
\expandafter\ifx\csname urlstyle\endcsname\relax
  \providecommand{\doi}[1]{doi: #1}\else
  \providecommand{\doi}{doi: \begingroup \urlstyle{rm}\Url}\fi

\bibitem[Ahn et~al.(2020)Ahn, Tucker, Wu, and Srivastava]{ahn-CSDA}
K.~Ahn, J.~D. Tucker, W.~Wu, and A.~Srivastava.
\newblock Regression models using shapes of functions as predictors.
\newblock \emph{Comp. Statistics \& Data Analysis}, 151:\penalty0 107017, 2020.

\bibitem[Ait-Sa{\"\i}di et~al.(2008)Ait-Sa{\"\i}di, Ferraty, Kassa, and
  Vieu]{AS2008}
A.~Ait-Sa{\"\i}di, F.~Ferraty, R.~Kassa, and P.~Vieu.
\newblock Cross-validated estimations in the single-functional index model.
\newblock \emph{Statistics}, 42\penalty0 (6):\penalty0 475--494, 2008.

\bibitem[Amato et~al.(2006)Amato, Antoniadis, and De~Feis]{A2006}
U.~Amato, A.~Antoniadis, and I.~De~Feis.
\newblock Dimension reduction in functional regression with applications.
\newblock \emph{Computational Statistics \& Data Analysis}, 50\penalty0
  (9):\penalty0 2422--2446, 2006.

\bibitem[Boj et~al.(2010)Boj, Delicado, and Fortiana]{boj2010distance}
E.~Boj, P.~Delicado, and J.~Fortiana.
\newblock Distance-based local linear regression for functional predictors.
\newblock \emph{Computational Statistics \& Data Analysis}, 54\penalty0
  (2):\penalty0 429--437, 2010.

\bibitem[Boj et~al.(2016)Boj, Caball{\'e}, Delicado, Esteve, and
  Fortiana]{boj2016global}
E.~Boj, A.~Caball{\'e}, P.~Delicado, A.~Esteve, and J.~Fortiana.
\newblock Global and local distance-based generalized linear models.
\newblock \emph{Test}, 25:\penalty0 170--195, 2016.

\bibitem[Delicado(2024)]{delicado2024comments}
P.~Delicado.
\newblock Comments on: Shape-based functional data analysis.
\newblock \emph{TEST}, 33\penalty0 (1):\penalty0 62--65, 2024.

\bibitem[Dryden and Mardia(2016)]{dryden2016statistical}
I.~L. Dryden and K.~V. Mardia.
\newblock \emph{Statistical Shape Analysis, with Applications in {R}. Second
  Edition.}
\newblock John Wiley and Sons, Chichester, 2016.

\bibitem[Du et~al.(2015)Du, Dryden, and Huang]{du2015size}
J.~Du, I.~L. Dryden, and X.~Huang.
\newblock Size and shape analysis of error-prone shape data.
\newblock \emph{Journal of the American Statistical Association}, 110\penalty0
  (509):\penalty0 368--379, 2015.

\bibitem[Eilers et~al.(2009)Eilers, Li, and Marx]{E2009}
P.~H.~C. Eilers, B.~Li, and B.~D. Marx.
\newblock Multivariate calibration with single-index signal regression.
\newblock \emph{Chemometrics and Intelligent Lab. Systems}, 96\penalty0
  (2):\penalty0 196--202, 2009.

\bibitem[Ferraty et~al.(2013)Ferraty, Goia, Salinelli, and Vieu]{F2013}
F.~Ferraty, A.~Goia, E.~Salinelli, and P.~Vieu.
\newblock Functional projection pursuit regression.
\newblock \emph{Test}, 22:\penalty0 293--320, 2013.

\bibitem[Ghosal et~al.(2023)Ghosal, Meiring, and Petersen]{ghosal2023}
A.~Ghosal, W.~Meiring, and A.~Petersen.
\newblock Fr{\'e}chet single index models for object response regression.
\newblock \emph{Electronic Journal of Statistics}, 17\penalty0 (1), 2023.

\bibitem[James and Silverman(2005)]{JS2005}
G.~M. James and B.~W. Silverman.
\newblock Functional adaptive model estimation.
\newblock \emph{Journal of the American Statistical Association}, 100\penalty0
  (470):\penalty0 565--576, 2005.

\bibitem[James et~al.(2009)James, Wang, and Zhu]{J2009}
G.~M. James, J.~Wang, and J.~Zhu.
\newblock Functional linear regression that's interpretable.
\newblock \emph{:}, pages 2083--2108, 2009.

\bibitem[Kendall et~al.(1999)Kendall, Barden, Carne, and
  Le]{kendall-barden-carne}
D.~G. Kendall, D.~Barden, T.~K. Carne, and H.~Le.
\newblock \emph{Shape and shape theory}.
\newblock Wiley, Chichester, New York, 1999.

\bibitem[Lee and Park(2012)]{LP2011}
E.~R. Lee and B.~U. Park.
\newblock Sparse estimation in functional linear regression.
\newblock \emph{Journal of Multivariate Analysis}, 105\penalty0 (1):\penalty0
  1--17, 2012.

\bibitem[Li et~al.(2010)Li, Wang, and Carroll]{YL2010}
Y.~Li, N.~Wang, and R.~J. Carroll.
\newblock Generalized functional linear models with semiparametric single-index
  interactions.
\newblock \emph{Journal of the American Statistical Association}, 105\penalty0
  (490):\penalty0 621--633, 2010.

\bibitem[Lin et~al.(2017)Lin, St.~Thomas, Zhu, and Dunson]{lin2017extrinsic}
L.~Lin, B.~St.~Thomas, H.~Zhu, and D.~B. Dunson.
\newblock Extrinsic local regression on manifold-valued data.
\newblock \emph{Journal of the American Statistical Association}, 112\penalty0
  (519):\penalty0 1261--1273, 2017.

\bibitem[Lin et~al.(2019)Lin, Mu, Cheung, and Dunson]{lin2019extrinsic}
L.~Lin, N.~Mu, P.~Cheung, and D.~Dunson.
\newblock Extrinsic gaussian processes for regression and classification on
  manifolds.
\newblock \emph{Bayesian Analysis}, 14\penalty0 (3), 2019.

\bibitem[Marron et~al.(2014)Marron, Ramsay, Sangalli, and
  Srivastava]{marron-etal-EJS:2014}
J.~S. Marron, J.~O. Ramsay, L.~M. Sangalli, and A.~Srivastava.
\newblock Statistics of time warpings and phase variations.
\newblock \emph{Electronic Journal of Statistics}, 8\penalty0 (2):\penalty0
  1697--1702, 2014.

\bibitem[Marron et~al.(2015)Marron, Ramsay, Sangalli, and
  Srivastava]{marron-etal-Statsc:2015}
J.~S. Marron, J.~O. Ramsay, L.~M. Sangalli, and A.~Srivastava.
\newblock {Functional Data Analysis of Amplitude and Phase Variation}.
\newblock \emph{Statistical Science}, 30\penalty0 (4):\penalty0 468 -- 484,
  2015.

\bibitem[Marx and Eilers(1999)]{ME1999}
B.~D. Marx and P.~H. Eilers.
\newblock Generalized linear regression on sampled signals and curves: a
  p-spline approach.
\newblock \emph{Technometrics}, 41\penalty0 (1):\penalty0 1--13, 1999.

\bibitem[McLean et~al.(2014)McLean, Hooker, Staicu, Scheipl, and
  Ruppert]{M2013}
M.~W. McLean, G.~Hooker, A.-M. Staicu, F.~Scheipl, and D.~Ruppert.
\newblock Functional generalized additive models.
\newblock \emph{Journal of Computational and Graphical Statistics}, 23\penalty0
  (1):\penalty0 249--269, 2014.

\bibitem[Morris(2015)]{JM2015}
J.~S. Morris.
\newblock Functional regression.
\newblock \emph{Annual Review of Statistics and Its Application}, 2:\penalty0
  321--359, 2015.

\bibitem[Petersen and M{\"u}ller(2019)]{petersen2019frechet}
A.~Petersen and H.-G. M{\"u}ller.
\newblock Fr{\'e}chet regression for random objects with euclidean predictors.
\newblock \emph{The Annals of Statistics}, 47\penalty0 (2):\penalty0 691--719,
  2019.

\bibitem[Ramsay and Silverman(2005)]{RS2005}
J.~O. Ramsay and B.~W. Silverman.
\newblock Fitting differential equations to functional data: Principal
  differential analysis.
\newblock \emph{Functional data analysis}, pages 327--348, 2005.

\bibitem[Randolph et~al.(2012)Randolph, Harezlak, and Feng]{R2012}
T.~W. Randolph, J.~Harezlak, and Z.~Feng.
\newblock Structured penalties for functional linear models—partially
  empirical eigenvectors for regression.
\newblock \emph{Electronic journal of statistics}, 6:\penalty0 323, 2012.

\bibitem[Reiss and Ogden(2007)]{RO2007}
P.~T. Reiss and R.~T. Ogden.
\newblock Functional principal component regression and functional partial
  least squares.
\newblock \emph{Journal of the American Statistical Association}, 102\penalty0
  (479):\penalty0 984--996, 2007.

\bibitem[Shi et~al.(2009)Shi, Styner, Lieberman, Ibrahim, Lin, and
  Zhu]{shi2009intrinsic}
X.~Shi, M.~Styner, J.~Lieberman, J.~G. Ibrahim, W.~Lin, and H.~Zhu.
\newblock Intrinsic regression models for manifold-valued data.
\newblock In \emph{International conference on medical image computing and
  computer-assisted intervention}, pages 192--199. Springer, 2009.

\bibitem[Shin and Oh(2022)]{shin2022robust}
H.-Y. Shin and H.-S. Oh.
\newblock Robust geodesic regression.
\newblock \emph{International Journal of Computer Vision}, 130\penalty0
  (2):\penalty0 478--503, 2022.

\bibitem[Srivastava and Klassen(2016)]{SK2016}
A.~Srivastava and E.~P. Klassen.
\newblock \emph{Functional and shape data analysis}, volume~1.
\newblock Springer, 2016.

\bibitem[Stoecker et~al.(2023)Stoecker, Steyer, and Greven]{stoecker-etal:2023}
A.~Stoecker, L.~Steyer, and S.~Greven.
\newblock Functional additive models on manifolds of planar shapes and forms.
\newblock \emph{Journal of Computational and Graphical Statistics}, 32\penalty0
  (4):\penalty0 1600--1612, 2023.

\bibitem[Thomas~Fletcher(2013)]{thomas2013geodesic}
P.~Thomas~Fletcher.
\newblock Geodesic regression and the theory of least squares on riemannian
  manifolds.
\newblock \emph{International journal of computer vision}, 105:\penalty0
  171--185, 2013.

\bibitem[Tsagkrasoulis and Montana(2018)]{tsagkrasoulis2018random}
D.~Tsagkrasoulis and G.~Montana.
\newblock Random forest regression for manifold-valued responses.
\newblock \emph{Pattern Recognition Letters}, 101:\penalty0 6--13, 2018.

\bibitem[Tucker et~al.(2013)Tucker, Wu, and Srivastava]{TUCKER201350}
J.~D. Tucker, W.~Wu, and A.~Srivastava.
\newblock Generative models for functional data using phase and amplitude
  separation.
\newblock \emph{Computational Statistics \& Data Analysis}, 61:\penalty0
  50--66, 2013.

\bibitem[Wu et~al.(2024)Wu, Huang, and Srivastava]{wu-TEST:2023}
Y.~Wu, C.~Huang, and A.~Srivastava.
\newblock Shape-based functional data analysis.
\newblock \emph{Test}, 33\penalty0 (1):\penalty0 1--47, 2024.

\bibitem[Zhang et~al.(2018)Zhang, Klassen, and Srivastava]{zhang-etal}
Z.~Zhang, E.~Klassen, and A.~Srivastava.
\newblock Phase-amplitude separation and modeling of spherical trajectories.
\newblock \emph{Journal of Computational and Graphical Statistics}, 27\penalty0
  (1):\penalty0 85--97, 2018.

\bibitem[Zhao et~al.(2012)Zhao, Ogden, and Reiss]{Z2012}
Y.~Zhao, R.~T. Ogden, and P.~T. Reiss.
\newblock Wavelet-based lasso in functional linear regression.
\newblock \emph{Journal of comp. and graphical statistics}, 21\penalty0
  (3):\penalty0 600--617, 2012.

\end{thebibliography}
\end{document}